\documentclass[twocolumn]{aastex62}
\usepackage{graphicx}
\usepackage{amssymb}
\usepackage{amsmath}
\usepackage{bm}
\usepackage{hyperref}
\usepackage{epstopdf}
\usepackage{color}

\graphicspath{{./}{figures/}}

\newcommand{\hammurabi}{\texttt{hammurabi}}
\newcommand{\hammurabiX}{\texttt{hammurabi\,X}}
\newcommand{\healpix}{\texttt{HEALPix}}
\newcommand{\fftw}{\texttt{FFTW}}
\newcommand{\cpp}{\texttt{C++}}

\newcommand{\namaster}{\texttt{NaMaster}}
\newcommand{\imagine}{\texttt{IMAGINE}}

\definecolor{burgundy}{rgb}{0.5, 0.0, 0.13}

\begin{document}
\title{hammurabi X: Simulating Galactic Synchrotron Emission with Random Magnetic Fields}

\correspondingauthor{Jiaxin Wang, Tess R. Jaffe,\\ Torsten A. En{\ss}lin, Piero Ullio}
\email{jiaxin.wang@sjtu.edu.cn,
tess.jaffe@nasa.gov,\\ ensslin@mpa-garching.mpg.de, piero.ullio@sissa.it}

\author[0000-0002-7384-7152]{Jiaxin Wang}
\affil{Scuola Internazionale Superiore di Studi Avanzati,
Via Bonomea 265, 34136 Trieste, Italy}
\affil{Department of Astronomy, Shanghai Jiao Tong University,
Shanghai, 200240, China}
\affil{Istituto Nazionale di Fisica Nucleare, Sezione di Trieste,
Via Bonomea 265, 34136 Trieste, Italy}

\author[0000-0003-2645-1339]{Tess R. Jaffe}
\affil{Department of Astronomy, University of Maryland,
College Park, MD, 20742, USA}
\affil{CRESST, NASA Goddard Space Flight Center,
Greenbelt, MD 20771, USA}

\author[0000-0001-5246-1624]{Torsten A. En{\ss}lin}
\affil{Max Planck Institute for Astrophysics,
Karl-Schwarzschild-Str. 1,
D-85741 Garching, Germany}

\author{Piero Ullio}
\affil{Scuola Internazionale Superiore di Studi Avanzati,
Via Bonomea 265, 34136 Trieste, Italy}
\affil{Istituto Nazionale di Fisica Nucleare, Sezione di Trieste,
Via Bonomea 265, 34136 Trieste, Italy}

\author[0000-0002-7546-0509]{Shamik Ghosh}
\affil{CAS Key Laboratory for Research in Galaxies and Cosmology, Department of Astronomy, University of Science and Technology of China, Hefei 230026, China}
\affil{School of Astronomy and Space Sciences, University of Science and Technology of China, Hefei, 230026,
China}

\author[0000-0001-7015-998X]{Larissa Santos}
\affil{Center for Gravitation and Cosmology, College of Physical Science and Technology, Yangzhou University,
Yangzhou 225009,
China}

\begin{abstract}
    We present version X of the \hammurabi\ package, the \healpix-based numeric simulator for Galactic polarized emission.
    Improving on its earlier design, we have fully renewed the framework with modern \cpp\ standards and features.
    Multi-threading support has been built in to meet the growing computational workload in future research.
    For the first time, we present precision profiles of \hammurabi\ line-of-sight integral kernel with multi-layer \healpix\ shells.
    In addition to fundamental improvements,
    this report focuses on simulating polarized synchrotron emission with Gaussian random magnetic fields.
    Two fast methods are proposed for realizing divergence-free random magnetic fields either on the Galactic scale where a field alignment and strength modulation are imposed, or on a local scale where more physically motivated models like a parameterized magneto-hydrodynamic (MHD) turbulence can be applied.
    As an example application, we discuss the phenomenological implications of Gaussian random magnetic fields for high Galactic latitude synchrotron foregrounds.
    In this, we numerically find B/E polarization mode ratios lower than unity based on Gaussian realizations of either MHD turbulent spectra or in spatially aligned magnetic fields.
\end{abstract}

\section{introduction}

The Galactic synchrotron emission from the diffuse distribution of relativistic electrons and positrons in the magnetized interstellar medium (ISM)\footnote{Acronyms used in the text:\\
CR (cosmic ray),\\
CMBR (cosmic microwave background radiation),\\
FFT (fast Fourier transform),\\
GMF (Galactic magnetic field),\\
ISM (interstellar medium),\\
LoS (line-of-sight),\\
MHD (magneto-hydrodynamics),\\
TE (thermal electron)
.}
is the dominant signal in the polarized sky observed at frequencies ranging from $\mathrm{MHz}$ to $\mathrm{GHz}$, therefore, it is one of the best friends to scientists who study multi-phase ISM structure and cosmic ray (CR) transport properties.  To those who study the cosmic microwave background radiation (CMBR), 21cm cosmology and the early Universe, however, it is one of their worst enemies.
Despite the difference between their scientific purposes, both fields recognize the importance of physical modelling of the mechanisms and environments associated with polarized synchrotron emission, absorption and Faraday rotation, which in the end provide a realistic description of the foreground observables.
The fundamental physical principles of the radiative transfer processes have been fully understood for around half a century \citep{Rybicki1979},
but with the growing precision and range of observations we are challenged by various local structures and non-linear phenomena within the Galaxy.
This is slowing down conceptual and theoretical advancements in related research fields since the observables are no longer analytically calculable in a high-resolution and non-perturbative regime.
To overcome the challenge, \hammurabi\ \citep{Waelkens2009} was developed to help us in simulating complicated observables with 3D modelling of the physical components of the Galaxy.

Over the last decade, we have witnessed wide scientific applications of \hammurabi\,
for example,
in estimating and removing Galactic synchrotron foreground contamination \citep{Dolag2015,Switzer2014},
in understanding magnetic fields of astrophysical objects varying from supernova remnants \citep{West2017} to the Galaxy \citep{Jaffe2013,Adam2016} and even to the local Universe \citep{Hutschenreuter2018}.
Despite the successful applications of \hammurabi\,, we have noticed that after years of modifications and the accumulation of modules and functions with outdated programming standards, the package might have been compromised by numeric issues and the lack of a properly maintained testing suite.
Given the trend towards high-resolution and computation-dominated studies, it is the right time to provide a precision guaranteed high-performance pipeline for simulating polarized synchrotron emission, absorption and Faraday rotation.
Thus a thorough upgrading project has been performed,
where we mainly focus on redesigning the code structure and work-flow, calibrating the numeric algorithms and methods, improving the user experience and setting up new conventions for future maintenance and development.

In addition to the technical improvements, we also keep up with recent progress in physical modelling of Galactic foreground emission with the turbulent Galactic magnetic field (GMF), e.g., phenomenological research carried out by \citet{Beck2016}, analytic estimations calculated by \citet{Cho2002,Caldwell2016,Kandel2017,Kandel2018},
and heavy simulations analyzed by \citet{Akahori2013,Kritsuk2018,Brandenburg2019}.
For future work about inferring the GMF configuration from observational data (e.g., Galactic synchrotron and dust emission, dispersion measure and Faraday rotation measure) we need physically motivated and numerically fast magnetic field simulators, instead of setting up trivial random fields or directly adopting expensive magneto-hydrodynamics (MHD) simulators.
The balance has to be made between the computational cost and the modelling complexity.
Low computational cost is required by any analysis that infers model parameters directly from data in a Bayesian fashion, where GMF models have to be evaluated repeatedly while the Bayesian inference algorithms sample through the often very high dimensional parameter space.
Full MHD simulations are currently prohibitively expensive to be used within such algorithms, there, fast emulators for the main statistical properties of typical MHD simulations are needed instead.

In this report, we propose two fast (in contrast to MHD simulation) random GMF generators which satisfy certain criteria.
A project for studying the GMF configuration with numeric simulation has been proposed \citep{imagine} by using a computational inference engine.
Though the main motivation for \hammurabiX\ is the construction of a Bayesian magnetic field inference engine, we herein present an analysis of the angular power spectrum focusing on the synchrotron B/E ratio as a possible guide for future studies.

This report is arranged as follows. 
In Section\,\ref{sec:hamx} we present a brief technical description of the \hammurabiX\ package with precision and performance profiles.
Section\,\ref{sec:gmf} presents mathematical details of the random GMF generators and the properties of their products.
In Section\,\ref{sec:powerspec}, we illustrate and discuss the influence of random GMF models on simulated synchrotron foreground angular power spectra.
A summary is provided as Section\,\ref{sec:sum} with prospects for future work.

Further more, in Appendix\,\ref{sec:sync_tech} we present the detailed numerical implementation of calculating the synchrotron emissivity and Faraday rotation. 
In Appendix\,\ref{sec:brnd_tech} we provide our method for vector field FFT in generating magnetic fields and its precision profile. 
The precision related to pseudo-$C_\ell$ estimation is addressed in Appendix\,\ref{sec:cl_tech}, and finally in Appendix\,\ref{sec:alternative_tech} we briefly discuss about the divergence cleaning in generating random magnetic fields.  

\section{hammurabi X}\label{sec:hamx}

\subsection{overview}

The \hammurabi\ code \citep{Waelkens2009} is an astrophysical simulator based on 3D models of the components of the magnetised ISM such as magnetic fields, thermal electrons, relativistic electrons, and dust grains.  
It performs an efficient line-of-sight (LoS) integral through the simulated Galaxy model using a \healpix-based\footnote{\url{https://healpix.jpl.nasa.gov}} \citep{Gorski2005} nested grid to produce observables such as Faraday rotation measure and diffuse synchrotron and thermal dust emission\footnote{This report focuses on the Galactic synchrotron emission, while the report for simulating thermal dust emission with \hammurabiX\ is under preparation.} in full Stokes $I$, $Q$ and $U$, while taking into account beam and depth depolarization as well as Faraday effects.
The updated version, \hammurabiX\ \citep{hamxjoss}\footnote{\hammurabiX\ is available in the repository \url{https://bitbucket.org/hammurabicode/hamx}, with detailed documentation. Recently, \hammurabiX\ has already been used to generate extra-Galactic Faraday rotation maps from primordial magnetic fields in \cite{Hutschenreuter2018}.}, has been developed to achieve higher computing performance and precision.
Specific efforts have been devoted to the parallel computing of LoS integral and vector filed FFT.

\hammurabiX\ currently uses the \healpix\ library \citep{Gorski2005} for observable production, where the LoS integral accumulates through several layers of spherical shells with adaptable \healpix\ resolutions.
We provide two modes of integral shell arrangements.
In the auto-shell mode, given $R$ as the maximum simulation radius, the $n^{\mathrm{th}}$ shell out of $N$ total shells covers the radial distance from $2^{(n-N-1)}R$ to $2^{(n-N)}R$, except for the first shell which starts at the observer.
The $n^{\mathrm{th}}$ shell is by default set up with the \healpix\ resolution controlling parameter $N_{\mathrm{side}} = 2^{(n-1)}N_{\mathrm{min}}$,\footnote{$N_\mathrm{side}$ means the number of full sky pixels is $12N^2_\mathrm{side}$.} where $N_{\mathrm{min}}$ represents the lowest simulation resolution at the first shell. 
Alternatively in the manual-shell mode, shells are defined explicitly by a series of dividing radii and \healpix\ $N_{\mathrm{side}}$'s.
The radial resolution along the LoS integral is uniformly set by the minimal radial distance for each shell.
The auto-shell mode follows the idea that the integral domain is discretized with elemental bins of the same volume, while the manual-shell mode allows users to refine specific regions to meet special realization requirements.

The LoS integral is carried out hierarchically:  at the top level the integral is divided into multiple shells with given spherical resolution settings, while at the bottom level inside each shell (where the spherical resolution is fixed) the radial integral is carried out with the midpoint rule for each radial bin.
Accumulation of observable information from the inner to outer shells is applied at the top level.
We emphasize that in \hammurabiX\,, the simulation spherical resolution for each shell can be independent of that in the outputs, which means that we can simulate with an arbitrary number of shells and assign each shell with a unique $N_\mathrm{side}$ value.
During each step of the shell accumulating process, we interpolate (with the linear interpolation provided by \healpix\ library) the current result into the output resolution.
Consequently, such interpolation between different angular resolutions will inevitably create a certain level of precision loss.

Previously in \hammurabi, the generation of the anisotropic component of the random field as well as the modulation of the field strength following various parametric forms lead to artificial magnetic field divergence.
Now we propose two improved solutions for simulating the random magnetic field.
On Galactic scales, a triple Fourier transform scheme is proposed to restore the divergence-free condition via a cleaning process.
This imposes the divergence-free property in the random magnetic field (unlike in \citealt{Adam2016}), which will be discussed in detail in Section\,\ref{sec:global_generator} with its observational implication in Section\,\ref{sec:powerspec}.
Alternatively, in a given local region\footnote{The local region means any small-scale spatial domain where the mean magnetic field can be treated or approximated as uniform distribution. This implies that the local generator cannot be applied to realize large-scale random magnetic fields, which are typically handled by the global generator.
In Section\,\ref{sec:powerspec} we will present and analyze local realizations at the solar neighbourhood as an example.},
a vector-field decomposition scheme is capable of simulating more detailed random field power-spectra.

\begin{figure}
    \plotone{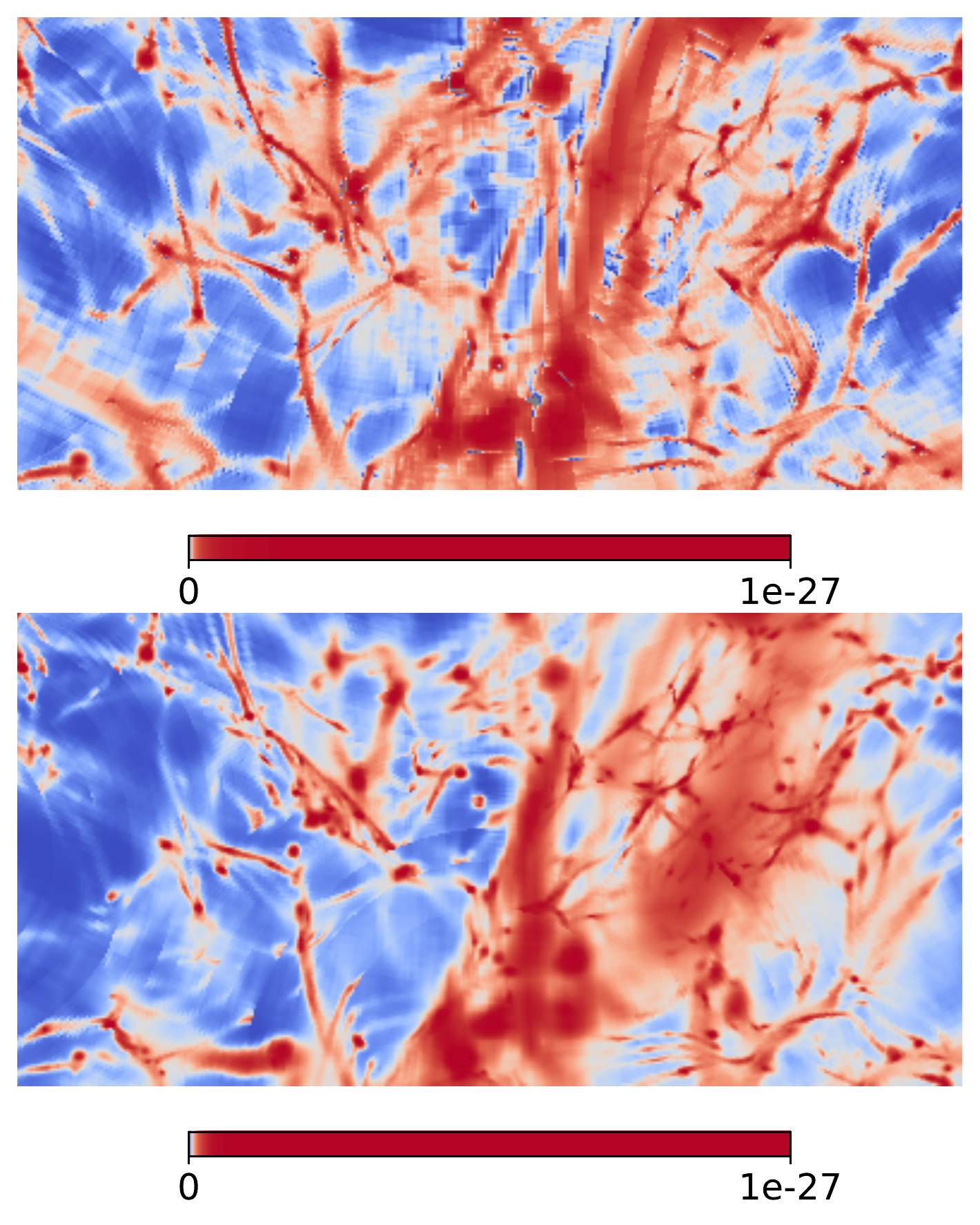}
    \caption{Comparison between the output from earlier version \hammurabi\ (top) and \hammurabiX\ (bottom).
    The sky patch in this illustration shows the extra-Galactic dispersion measure (an observable with non-negative value by definition) simulated and studied by \citet{Hutschenreuter2018}.}\label{fig:comparison}
\end{figure}

Fast Fourier transforms (FFTs) are necessary for translating the power spectra of random fields into discrete magnetic field realizations on 3D spatial grids.
Random field generators in \hammurabiX\ currently use the \fftw\ library\footnote{\url{http://www.fftw.org}}.
The detailed implementation will be discussed in Section\,\ref{sec:gmf}.
In cases where the field is input from an external or internal discrete grid, e.g., a random GMF, the LoS integral at a given position does linear interpolation (in each phase-space dimension) from nearby grid points.
The interpolation algorithm has been calibrated, so the high-resolution outputs are no longer contaminated by the numerical flaws in earlier versions of \hammurabi.
As illustrated in Figure\,\ref{fig:comparison}, the interpolation process in the earlier version of \hammurabi\ did not properly calculate the volume of elemental discretization, which resulted, for example, in negative values of the simulated dispersion measure and incorrect small scale features in comparison to the corrected method in \hammurabiX.
In this new version, unit tests for linear interpolation can be found in the public repository.

Generally speaking, the precision of the linear interpolation (and the corresponding discretization) can in principle be characterized by the goodness of the approximation. 
This is explicitly affected by the discretization resolution and the arrangement of the sampling/supporting points, and also by the smoothness (as measured by the inverse of the second order derivative) of the approximation target.
In \hammurabiX\ the interpolation affects the precision in realizing the power spectrum of the random magnetic field generation. 
This can be improved by increasing the sampling resolution.
Furthermore, the linear interpolation does not preserve the divergence, but the precision can be improved either by increasing sampling resolution\footnote{If we estimate the divergence by the finite difference in the spatial domain, the precision exponentially improves as a function of the number of sample points in each direction.} or by matching the elemental discretization volume in the LoS integral with that in the field generation (as discussed by \citet{Waelkens2009}).

\subsection{precision and performance profiles}

Profiling\footnote{The \hammurabiX\ wiki page \url{https://bitbucket.org/hammurabicode/hamx/wiki/Home} presents detailed verification, performance and precision profiles, implementation methods and online documentation.} the numerical precision in producing observables is critical in guiding practical applications.
A standard \hammurabiX simulation routine consists of two major building blocks.
The first part is the numerical implementation of specific physical processes like synchrotron emission and Faraday rotation, and the second part is the LoS integral that is universal to all observables.
In the following integrated precision check, the correctness of both will be verified and profiled together.

A given magnetic field vector $\mathbf{B}$ can be decomposed into directions parallel (horizontal) and perpendicular (vertical/poloidal) to the Galactic disk, or to be specific, the $\{\hat{\mathbf{x}},\hat{\mathbf{y}}\}$ plane (with $\hat{\mathbf{y}}$ pointing towards Galactic longitude $l=90^\circ$) in the \hammurabiX\ convention,
i.e., $\mathbf{B}_\parallel$ and $\mathbf{B}_\perp$ at a given Galactic longitude-latitude position $\{l,b\}$.
The LoS direction $\hat{\mathbf{n}}$ from the observer to the target field position reads
\begin{eqnarray}
    \hat{\mathbf{n}} &=& \cos(b)\cos(l)\hat{\mathbf{x}} + \cos(b)\sin(l)\hat{\mathbf{y}} + \sin(b)\hat{\mathbf{z}} ~,
\end{eqnarray}
where $\hat{\mathbf{x}}$ is conventionally pointing from the observer to the Galactic centre. 
In the same observer-centric Cartesian frame we can explicitly write down two field components as
\begin{eqnarray}
    \mathbf{B}_\parallel &=& B_\parallel( \cos(l_0)\hat{\mathbf{x}} + \sin(l_0)\hat{\mathbf{y}} ) ~,\\
    \mathbf{B}_\perp &=& B_\perp \hat{\mathbf{z}} ~,
\end{eqnarray}
where $l_0$ represents the projected direction of $\mathbf{B}$ in the $\{\hat{\mathbf{x}},\hat{\mathbf{y}}\}$ plane.
Then it is straight forward to calculate two key quantities for the calculation of synchrotron emissivity and Faraday rotation respectively
\begin{eqnarray}
    |\mathbf{B}\times\hat{\mathbf{n}}| &=& \sqrt{B^2_\parallel+B^2_\perp-|\mathbf{B}\cdot\hat{\mathbf{n}}|^2} ~, \\
    \mathbf{B}\cdot\hat{\mathbf{n}} &=& B_\parallel\cos(b)\cos(l-l_0) + B_\perp\sin(b) ~,
\end{eqnarray}
It is obvious that Faraday rotation is more sensitive to $\mathbf{B}_\parallel$ at low Galactic latitudes, and to $\mathbf{B}_\perp$ at high latitudes. 
On the contrary, synchrotron emissivity, which is proportional to some power of $|\mathbf{B}\times\hat{\mathbf{n}}|$, 
is more sensitive to $\mathbf{B}_\perp$ at low Galactic latitudes and to $\mathbf{B}_\parallel$ at high latitudes.

\begin{figure}
    \plotone{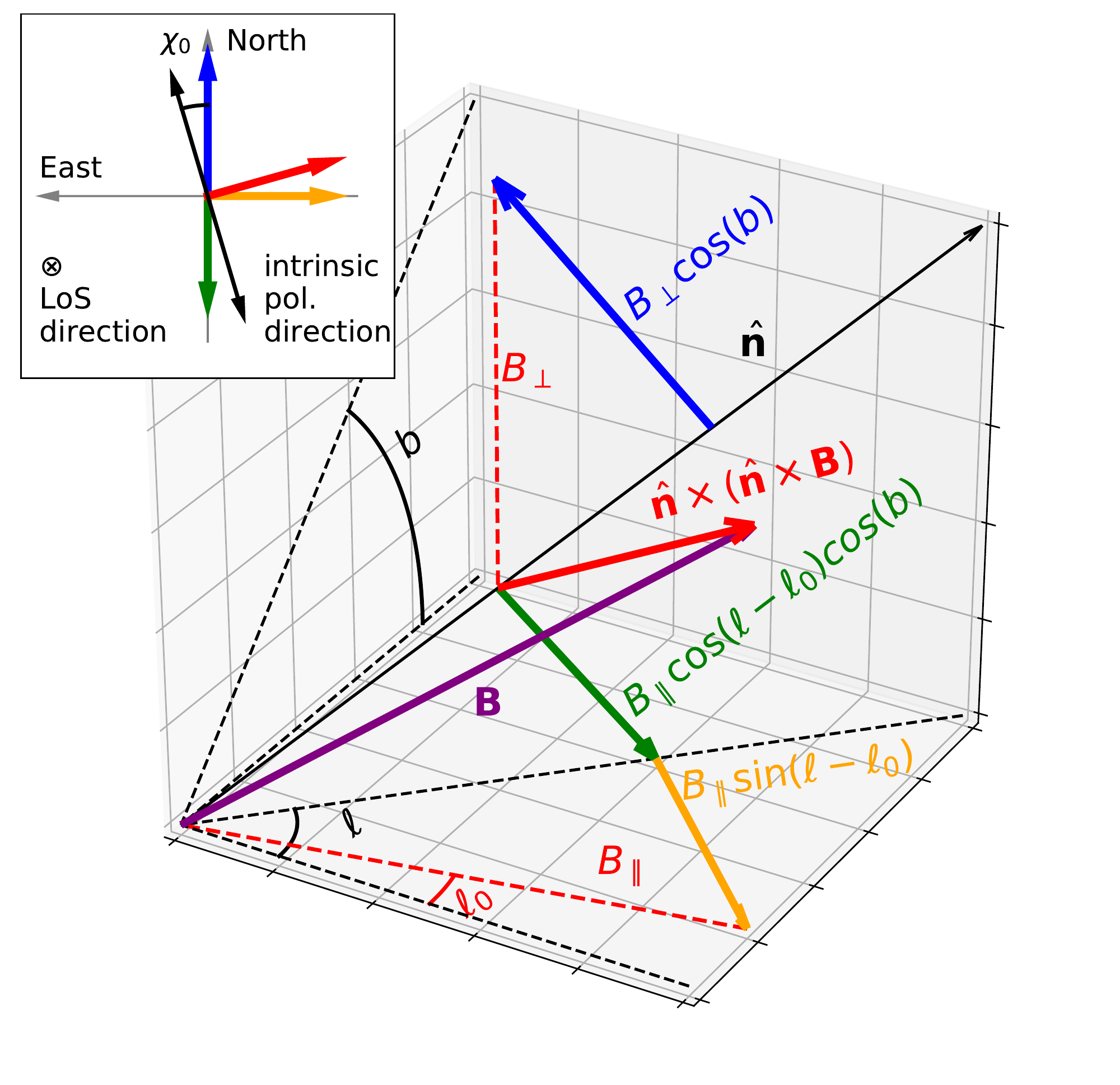}
    \caption{Cartoon illustration of the projection of magnetic field $\mathbf{B}$ to the LoS direction $\hat{\mathbf{n}}$.
    The definition of synchrotron intrinsic polarization angle (with north-to-east as the positive angle direction) is presented on the top left, the plan of the sky, with red arrow presenting the magnetic field projected to it.}\label{fig:b_decompose}
\end{figure}

Precision checks require a baseline model for each field, from which analytic descriptions of the observables can be explicitly derived.
Here we assume spatially homogeneous distributions for the cosmic-ray electrons (CREs), thermal electrons (TEs) and the GMF within a given radial distance to observer. 
The spectral index of the CRE energy distribution is assumed to be a constant, and consequently CRE density $N(\gamma)$ is described by
\begin{eqnarray}
    N(\gamma) &=& N_0\gamma^{-\alpha}~,
\end{eqnarray}
where $\gamma$ represents CRE Lorentz factor, $\alpha$ represents the constant spectral index of CRE.
With the assumed homogeneity in all fields, we can calculate intrinsic synchrotron total intensity $I_0$ and polarization Stokes parameter $Q_0$ and $U_0$ (in the IAU convention\footnote{Detailed description for IAU and CMB polarization conventions can be found at \url{https://lambda.gsfc.nasa.gov/product/about/pol_convention.cfm}.}) before applying the Faraday rotation \citep{Rybicki1979}
\begin{eqnarray}
    I_0 &=& J_{i}R_0 ~,\\ 
    Q_0 &=& J_\mathrm{pi}R_0\cos(2\chi_0) ~,\\
    U_0 &=& J_\mathrm{pi}R_0\sin(2\chi_0)~, \\
    J_i &=& \frac{\sqrt{3}e^3|\mathbf{B}\times\hat{\mathbf{n}}|N_0}{4\pi m_\mathrm{e}c^2(\alpha+1)} \left(\frac{2\pi\nu m_\mathrm{e}c}{3e|\mathbf{B}\times\hat{\mathbf{n}}|}\right)^{\frac{1-\alpha}{2}} \\\nonumber
    &&\times\Gamma\left(\frac{\alpha}{4}+\frac{19}{12}\right)\Gamma\left(\frac{\alpha}{4}-\frac{1}{12}\right)~,\\
    J_{pi} &=& \frac{\sqrt{3}e^3|\mathbf{B}\times\hat{\mathbf{n}}|N_0}{16\pi m_\mathrm{e}c^2} \left(\frac{2\pi\nu m_\mathrm{e}c}{3e|\mathbf{B}\times\hat{\mathbf{n}}|}\right)^{\frac{1-\alpha}{2}} \\\nonumber
    &&\times\Gamma\left(\frac{\alpha}{4}+\frac{7}{12}\right)\Gamma\left(\frac{\alpha}{4}-\frac{1}{12}\right) ~,
\end{eqnarray}
where $e$ is the electron charge, and $m_\mathrm{e}$ is the electron mass, $R_0$ is the spherical LoS integral depth, and $\nu$ is the observational frequency. 
The intrinsic polarization angle $\chi_0$ can be derived from
\begin{eqnarray}
    \tan(\chi_0) &=& \frac{B_\perp\cos(b)-B_\parallel\sin(b)\cos(l-l_0)}{B_\parallel\sin(l-l_0)} ~,
\end{eqnarray}
as illustrated in Figure\,\ref{fig:b_decompose}.
With the same modelling, Faraday depth $\phi$ can be described by
\begin{eqnarray} \label{eq:fd_ana}
    \phi(l,b) &=& \phi_0 R_0 ~,\\
    \phi_0 &=& -N_\mathrm{e}(\mathbf{B}\cdot\hat{\mathbf{n}})\left(\frac{e^3}{2\pi m_\mathrm{e}^2c^4}\right) ~,
\end{eqnarray}
where $N_\mathrm{e}$ represents constant homogeneous TE density assumed within spherical radius $R_0$.
In the end, the observed synchrotron polarization Stokes parameters $Q$ and $U$ reflect the Faraday rotation as
\begin{eqnarray}
    Q+iU = (Q_0+iU_0)\int^{R_0}_0 \frac{e^{2i\phi_0\lambda^2 r}}{R_0} dr ~,
\end{eqnarray}
which indicates that the polarized intensity receive a correction factor $|\sin(\phi\lambda^2)/(\phi\lambda^2)|$ known as the Faraday depolarization.
The formulae above analytically derive calculable results for reference in verifying the numerical outputs.
In real applications, the magnetic field and CRE spectral index are not constant, and the methods used by \hammurabiX\ for calculating synchrotron emissivity and Faraday rotation can be more generic, as presented in Appendix\,\ref{sec:sync_tech}.

\begin{figure}
    \plotone{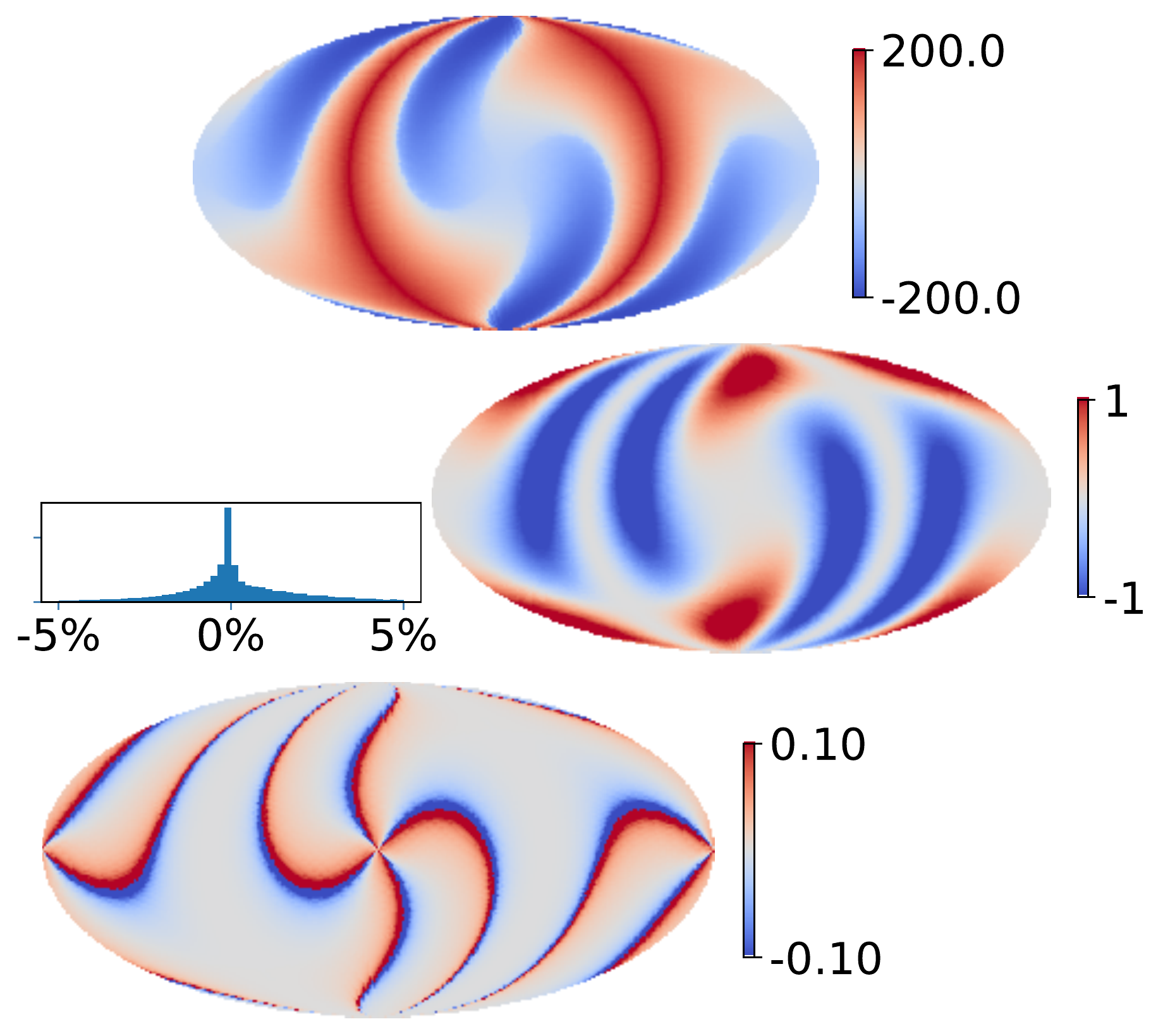}
    \caption{Synchrotron Stokes Q (top) at $2.4~\mathrm{GHz}$ where the influence of Faraday rotation is clearly imprinted.
    Absolute error (middle) and relative error (bottom) are presented according to the analytic reference with $B_\perp = 0$ and $l_0 = 0$.
    The histogram (middle left) presents relative error distribution.
    The single shell LoS integral is carried out with radial resolution set as $1\%$ of the total radius.}
\label{fig:err_spin2}
\end{figure}

Figure\,\ref{fig:err_spin0} presents the absolute and relative numeric error distribution of synchrotron total intensity from a single LoS integral shell.
For an observable $X$, the absolute error is defined as the difference between simulated output $X_\mathrm{sim}$ and the analytic reference $X_\mathrm{ref}$ as $(X_\mathrm{sim}-X_\mathrm{ref})$, while the relative error is defined by $2(X_\mathrm{sim}-X_\mathrm{ref})/(X_\mathrm{sim}+X_\mathrm{ref})$.
The Faraday depth calculator shares a similar error distribution as the calculator of synchrotron total intensity.
Meanwhile, Figure\,\ref{fig:err_spin2} presents the absolute and relative numeric error distributions of synchrotron Stokes $Q$ also from a single LoS integral shell, which serves as an example for illustrating the numeric precision in calculating tensor fields.
With constant field models in testing, the numeric errors are mainly induced by the integration and interpolation methods and therefore independent of the LoS resolution.
Even with simple field settings, we can observe a few percent relative error appearing in Figure\,\ref{fig:err_spin2}.
Considering the future usage of \hammurabiX\ in inferring Galactic component structures with astrophysical measurements, if the magnitude of such numerical errors are larger than the observational uncertainties, a Bayesian analysis with \hammurabiX\ will consequently suffer from higher uncertainties and bias in parameter estimation.

\begin{figure}
    \plotone{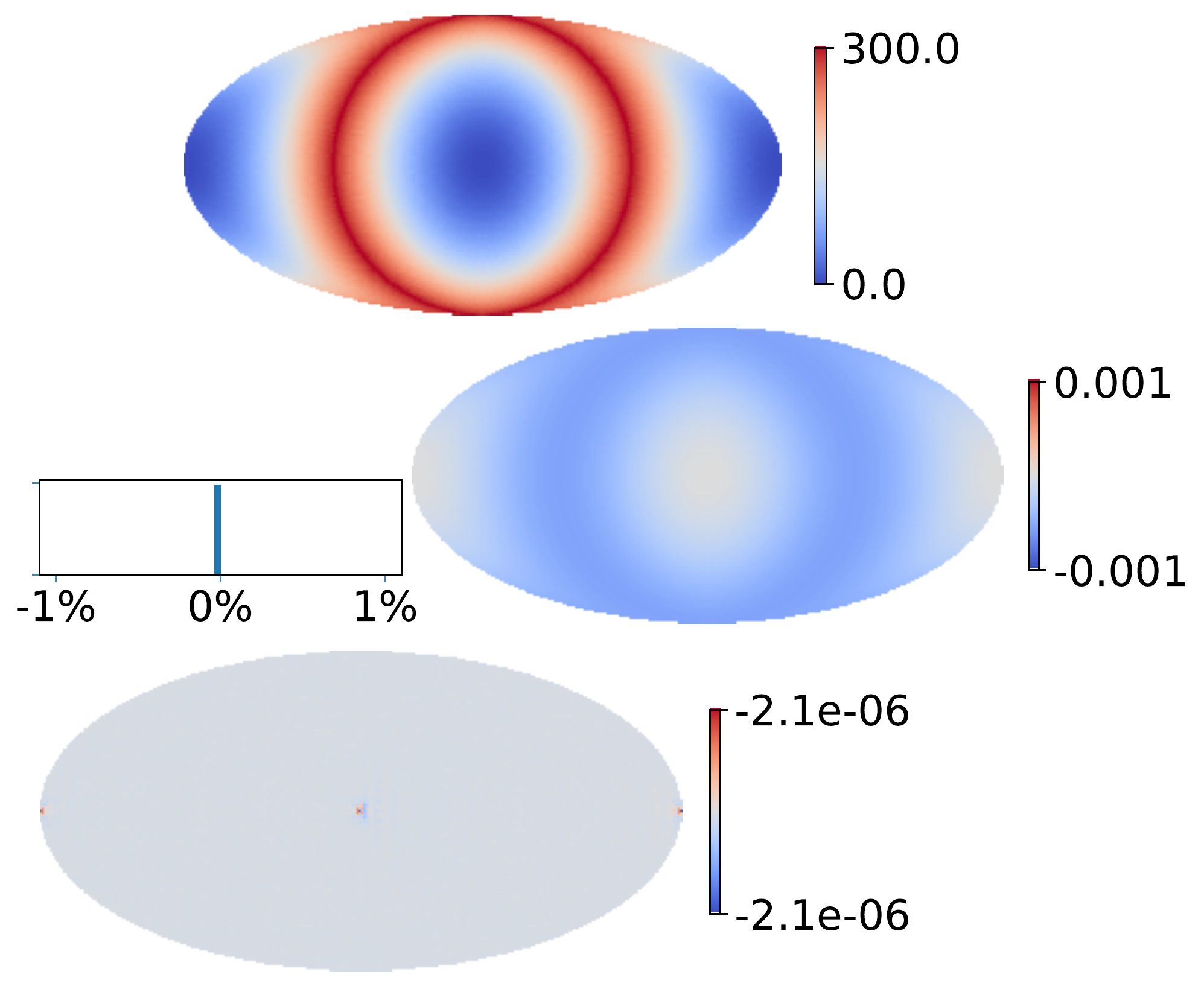}
    \caption{Synchrotron Stokes I (top) at $2.4~\mathrm{GHz}$. 
    Absolute error (middle) and relative error (bottom) are presented according to the analytic reference with $B_\perp = 0$ and $l_0 = 0$.
    The histogram (middle left) presents relative error distribution.
    The single shell LoS integral is carried out with radial resolution set as $1\%$ of the total radius.}
\label{fig:err_spin0}
\end{figure}

In terms of the multi-shell arrangement in real application, the output precision is affected by the spherical surface interpolation provided by the \healpix\ library.
The motivation of allowing different resolution settings along with the divided LoS integral shells is to save computing resources as mentioned in \citet{Adam2016}.
It is worth noticing that in the simulation, the pixel values are calculated along their central spherical coordinates. 
This is different from the actual astrophysical measurements where each pixel value is estimated based on many observational hits.
And thus for quickly comparing low resolution simulation results with high-resolution data, we recommend interpolating data on the sky, concerning simulations' sample directions, instead of downgrading data by averaging over high-resolution pixels.
In this way, we avoid comparing exactly predicted values of simulation to region-averaged values of measurements.
Alternatively, a very stringent simulation should be designed to mimic the true observation beams, which is computational heavy without \hammurabi's method.
But even with our method, no simulation can capture reality perfectly, and the user must always be careful to test that the simulation resolution is sufficient for probing the observational property in question.  

The testing cases displayed above are prepared by assuming constant magnetic field and thermal electron field distributions.
The numerical errors would inevitably grow larger when the input Galactic components have small scale features near or below the discretization resolution.
This issue can be handled efficiently in the future by an adaptively refined mesh/pixelization.

\begin{figure}
    \plotone{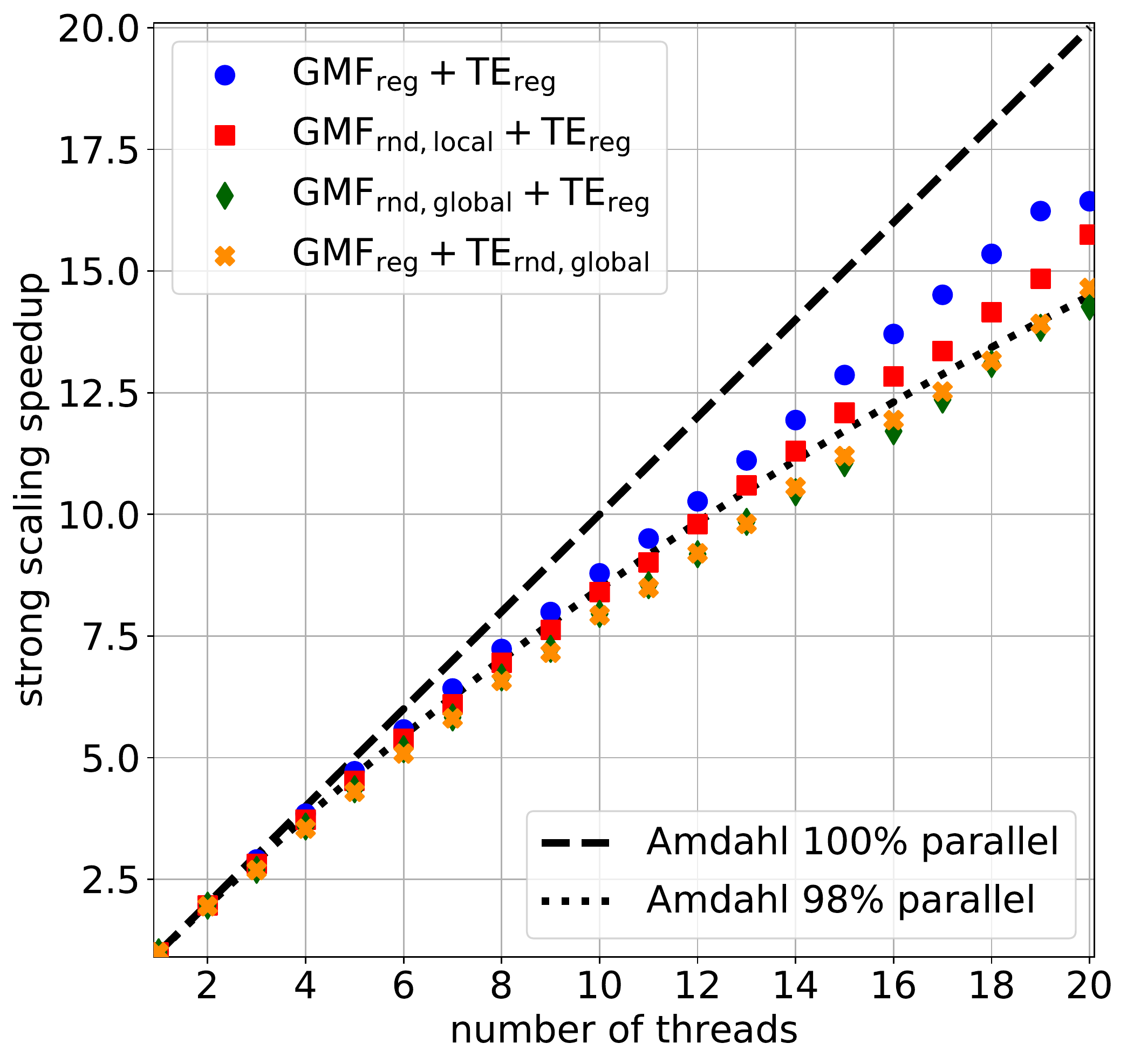}
    \caption{\hammurabiX\ strong scaling speedups in various tasks, where the subscript ``reg'' stands for regular fields while ``rnd'' stands for random fields.
    No bottle neck from memory access has been observed.
    The simulation routines are set by default as calculating synchrotron emission with Faraday rotation.}\label{fig:hamx_speedup}
\end{figure}

The computationally heavy processes in \hammurabiX\ are the LoS integration for \healpix\ map pixels, 
the random field generation with FFTs,
and the linear interpolation for fields prepared in grids (e.g., internal random fields and other external fields).
Massive observable production, \healpix\ map distribution and recycling of physical fields require MPI\footnote{Message Passing Interface (MPI) is a standardized and portable message-passing standard designed by a group of researchers from academia and industry to function on a wide variety of parallel computing architectures.} parallelization and therefore are beyond our scope in this report.
In this work, multi-threading is always essential at the bottom level of parallelism.
Figure\,\ref{fig:hamx_speedup} presents the strong scaling\footnote{Strong scaling is defined as how the solution time varies with the number of processors for a fixed total problem size.} in observable production with various GMF and TE field combinations.
The strong scaling with either computationally heavy (with random field generation) or light (without random field generation) pipelines follows the Amdahl law \citep{Amdahl1967} with around $2\%$ serial remnants.
Note that the speedup properties are not very sensitive to the resolution setting in various simulation routines, since the workload of pure numerical operations is proportional to the discretization resolution.

\section{Gaussian random GMF}\label{sec:gmf}

\subsection{general discussion}

Realization of turbulent magnetic field is a major module in \hammurabiX,
since the correctness of most simulations relies on physically motivated and accurate description of the turbulent fields in the multi-phase ISM.
In this section we present two Gaussian random GMF  generators that are by definition divergence-free and capable of realizing field alignment and/or strength modulation on Galactic scales or an anisotropic\footnote{In this work, spatially anisotropic random GMF means it is locally aligned either parallel or perpendicular to a preferred direction (e.g., by alignment parameter $\rho$ in the global random GMF generator), while spectral anisotropy means the anisotropy in the frequency domain (usually due to an anisotropic power spectrum, e.g., the MHD turbulent magnetic field). We emphasize that in a local MHD turbulent magnetic field realization, the spectral anisotropy results in the spatially anisotropic distribution.} power spectrum on small scales.

There are several criteria that a random GMF generator should satisfy.
That it be divergence-free (or solenoidal) is always the prime feature of any magnetic field.
Absolute zero divergence is hard to define under discretisation, but in principle either a vector-field decomposition or a Gram-Schmidt process in the frequency domain is capable of cleaning field divergence.
In realistic cases  when a large-scale spatial domain is expected to be filled with random magnetic fields, the field strength and alignment need to be correlated with the large-scale structures in the Galaxy.
This requirement complicates the generating process, because the divergence-free property should also be satisfied simultaneously. It is straightforward to generate a divergence-free Gaussian random field, and equally simple to then re-scale or stretch it as done in \citet{Jaffe2010}.  But the re-profiling process destroys the divergence-free property if it is applied naively.

A triple Fourier transform scheme is thus proposed mainly to reconcile these two requirements.
At Galactic scales, the new scheme allows modification of the Gaussian random realization by a given inhomogeneous spatial profile for the field strength.
Note that aligning the magnetic field to a given direction is easy to implement in the spatial domain,
but locally varying anisotropy in the energy power spectra is not feasible by a single FFT.
In studies of Galactic emission from MHD plasma, the dependency of local structure on a varying direction profile breaks the symmetry required for using the FFT.
To perform more detailed modelling of the turbulent GMF power spectrum, we provide a local generator (`local' in the sense that the mean field can be approximated in uniform distribution) with explicit or implicit vector decomposition.

\subsection{power spectrum}

Consider a magnetic field distribution $\mathbf{B(x)} = \mathbf{B}_0(\mathbf{x}) + \mathbf{b(x)}$ and its counterpart $\mathbf{\tilde{B}(k)}$ in the frequency domain, where $\mathbf{B}_0$ and $\mathbf{b}$ represent the regular and turbulent fields respectively.
The simplest turbulent power spectrum is represented by the trace of the isotropic magnetic field spectrum tensor in scalar form, $P(k) \propto \langle \mathbf{\tilde{B}(k)}\cdot\mathbf{\tilde{B}^\ast(k)} \rangle_\mathbf{B}$\footnote{$\langle...\rangle_\mathbf{B}$ means an ensemble average over all $\mathbf{B}$ field realizations.}.
This kind of spectrum is widely used as a first approach to the turbulent field realization where the spectral shape is important.
In general we could parameterize the basic scalar spectrum as
\begin{eqnarray}\label{eq:magnetic_spectrum}
    P(k) &=& \frac{P_0}{4\pi k^2} \Big[
    \left(\frac{k_0}{k_1}\right)^{\alpha_1} \left(\frac{k}{k_1}\right)^6\mathcal{H}(k_1-k) \\\nonumber
    && + \left(\frac{k}{k_0}\right)^{-\alpha_1}\mathcal{H}(k-k_1)\mathcal{H}(k_0-k) \\\nonumber
    && + \left(\frac{k}{k_0}\right)^{-\alpha_0}\mathcal{H}(k-k_0) \Big]~,
\end{eqnarray}
where $\mathcal{H}$ represents the Heaviside step function.
The last term in equation\,\eqref{eq:magnetic_spectrum} represents the forward magnetic cascading of MHD turbulence from the injection scale $k_0$ to small scales ($k>k_0$),
while the first two terms describe the inverse cascading \citep{Pouquet1976} in MHD turbulence from $k_0$ to scale $k_1 \simeq 1/L$ which corresponds to the physical size $L$ of the MHD system.
According to the simulation results from \cite{Brandenburg2019}, we set $k_1 = 0.1~\mathrm{kpc^{-1}}$ and $\alpha_1=0.0$ by default in this work if not specified.
Note that although not explicitly mentioned here, the Nyquist frequency cutoff $k_\mathrm{max}$ requires an extra Heaviside factor $\mathcal{H}(k_\mathrm{max}-k)$ in equation\,\eqref{eq:magnetic_spectrum}.

In terms of more physical parameterization, we are interested in realizing theoretical descriptions of turbulence in the compressible plasma recently discussed by \citet{Cho2002}, \citet{Caldwell2016} and \citet{Kandel2017}.
In a compressible plasma, turbulence can be decomposed into Alfv{\'e}n, fast and slow modes.
Two critical plasma status parameters are the ratio $\beta$ and the Alfv{\'e}n Mach number $M_\mathrm{A}$.
The plasma $\beta$ is the ratio of gas pressure to magnetic pressure, which represents compressibility of the plasma, with $\beta \rightarrow \infty$ indicating the in-compressible regime.
The Alfv{\'e}n Mach number is the ratio of the injection velocity to the Alfv{\'e}n velocity, with $M_\mathrm{A} > 1.0$ representing the super-Alfv{\'e}nic regime while $M_\mathrm{A} < 1.0$ means sub-Alfv{\'e}nic turbulence.
The general form of the compressible MHD magnetic field spectrum tensor trace reads
\begin{eqnarray}\label{eq:scalarspec}
    P(k,\alpha) &=& \sum_i P_i(k)F_i(M_\mathrm{A},\alpha)h_i(\beta,\alpha) ~,
\end{eqnarray}
where $i = \{\mathrm{A},\mathrm{f},\mathrm{s}\}$ denotes one of the three MHD modes as described in detail in Section\,\ref{sec:mhd}.
In \hammurabiX, compressible MHD is only realized by the local generator, thus $\cos(\alpha) = \mathbf{\hat{k}\cdot\hat{B}_0}$ is adopted with $\mathbf{B}_0$ taken as the regular field near the observer.
A detailed application example of $F_i$ and $h_i$ is presented in Section\,\ref{sec:powerspec}.
Some additional information can be found in Appendix\,\ref{sec:brnd_tech} for readers who are interested in the technical shortcuts in random field generation and the sampling precision.

\subsection{global random GMF generator}\label{sec:global_generator}

One major task of \hammurabiX\ is to generate a random GMF that can cover a specific scale in the spatial domain. 
However, an inhomogeneous correlation structure is not diagonal in the frequency domain.
In this case, we try to impose an energy density and alignment profile in the spatial domain after the random realization is generated in the frequency domain with an isotropic spectrum. 
Then the field divergence can be cleaned back in the frequency domain with the Gram-Schmidt process.
The whole procedure of this scheme requires two backward and one forward FFTs.

After a Gaussian random magnetic field is realized in the frequency domain, 
each grid point holds a vector $\mathbf{b}$ drawn from an isotropic field dispersion. 
The key to the triple transform is the large-scale alignment and energy density modulation process. 
The alignment direction $\hat{\mathbf{H}}$ at different Galactic positions should be pre-defined like the energy density profile. 
We introduce the alignment parameter $\rho$ for imposing the alignment profile by
\begin{eqnarray} \label{eq:order_rpf}
    \mathbf{b(x)} &\rightarrow& \frac{(\mathbf{b}_\parallel\rho+\mathbf{b}_\perp/\rho)}{\sqrt{\frac{1}{3}\rho^2 + \frac{2}{3}\rho^{-2}}} ~,\\
    \mathbf{b}_\parallel &=& \frac{(\mathbf{b}\cdot\hat{\mathbf{H}})}{|\hat{\mathbf{H}}|^2}\hat{\mathbf{H}} ~,\\
    \mathbf{b}_\perp &=& \frac{\hat{\mathbf{H}}\times(\mathbf{b}\times\hat{\mathbf{H}})}{|\hat{\mathbf{H}}|^2}.
\end{eqnarray}
$\rho=1.0$ means no preferred alignment direction, while $\rho \rightarrow 0$ ($\rho \rightarrow \infty$) indicates extremely perpendicular (parallel) alignment with respect to $\hat{\mathbf{H}}$.
(Previously, the alignment operation in \hammurabi\ was carried out by regulating $\mathbf{b}_\parallel$ only \citep{Jaffe2010},
which is phenomenological equivalent to our approach presented here.) 
Note that $\rho$ and $\hat{\mathbf{H}}$ can either be defined as a global constant or as a function of other physical quantities such as the regular magnetic field and the Galactic ISM structure (detailed description can be found in the \hammurabiX\ wiki page).

For regulating the field energy density, a simple example with exponential scaling profile (which can be customized in future studies) is proposed as
\begin{eqnarray}
    S(\mathbf{x}) &=& \exp\left(\frac{R_\odot-r}{h_r}\right)\exp\left(\frac{|z_\odot|-|z|}{h_z}\right) ~,
\end{eqnarray}
where $(r,z)$ is the coordinate in the Galactic cylindrical frame, and $(R_\odot,z_\odot)$ represents the solar position in the Galactic cylindrical frame.
The energy density modulation acts on the vector field amplitude through 
\begin{eqnarray}\label{eq:energy_rpf}
    \mathbf{b(x)} \rightarrow \mathbf{b(x)}\sqrt{S(\mathbf{x})} ~.
\end{eqnarray}

The above operations of reorienting, stretching and squeezing magnetic field vectors in the spatial domain do not promise a divergence-free result.
To clean the divergence, we transform the re-profiled field forward into the frequency domain and apply the Gram-Schmidt process
\begin{eqnarray}\label{eq:gs}
    \mathbf{\tilde{b}} \rightarrow \sqrt{3}\left(\mathbf{\tilde{b}} - \frac{\mathbf{(k\cdot\tilde{b})k}}{|k|^2}\right) ~,
\end{eqnarray}
where $\mathbf{\tilde{b}}$ indicates the frequency-domain complex vector.
The coefficient $\sqrt{3}$ is for preserving the spectral power statistically.
The second backward Fourier transform is then carried out to provide the final random GMF vector distribution in the spatial domain.

\begin{figure}
    \plotone{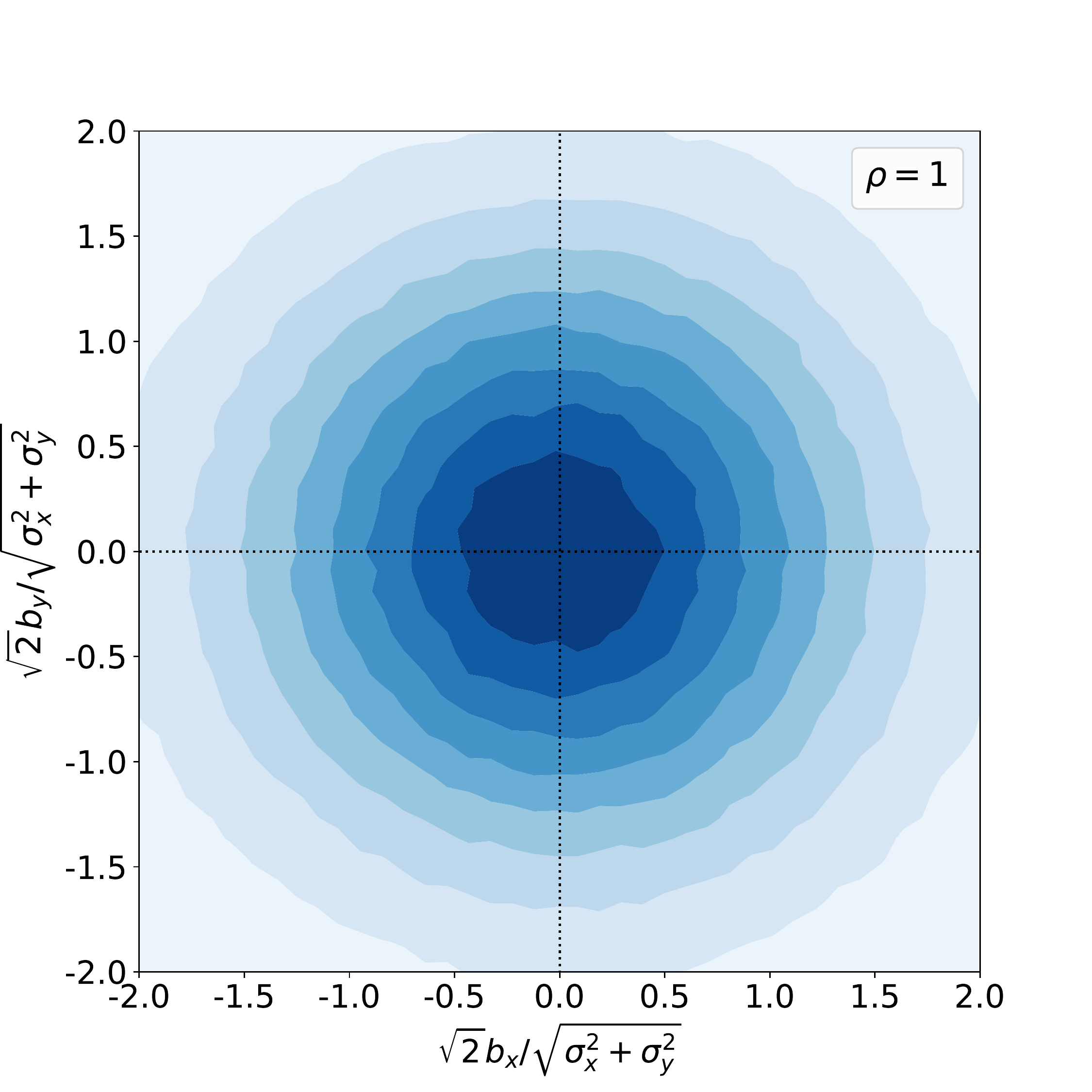}
    \plotone{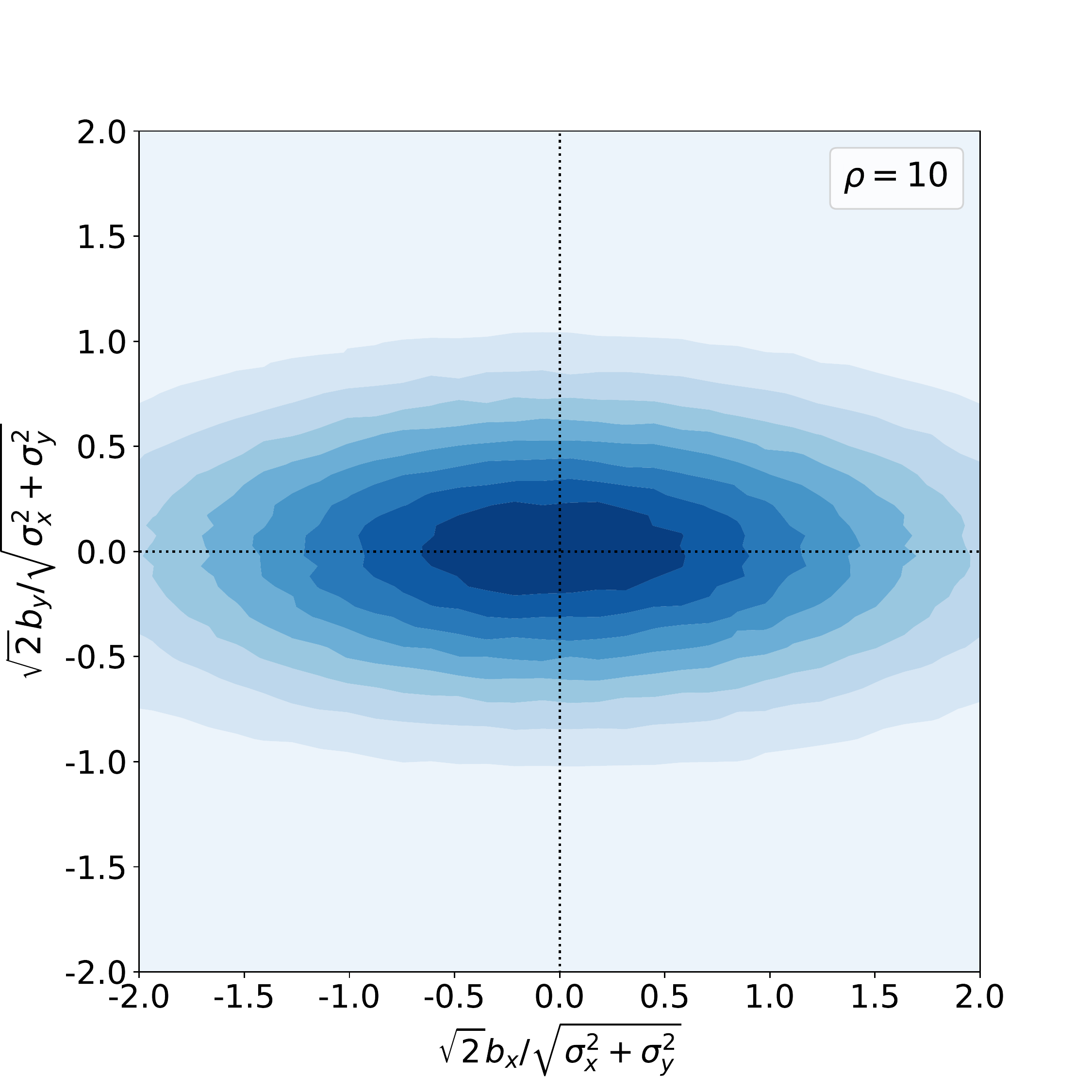}
    \plotone{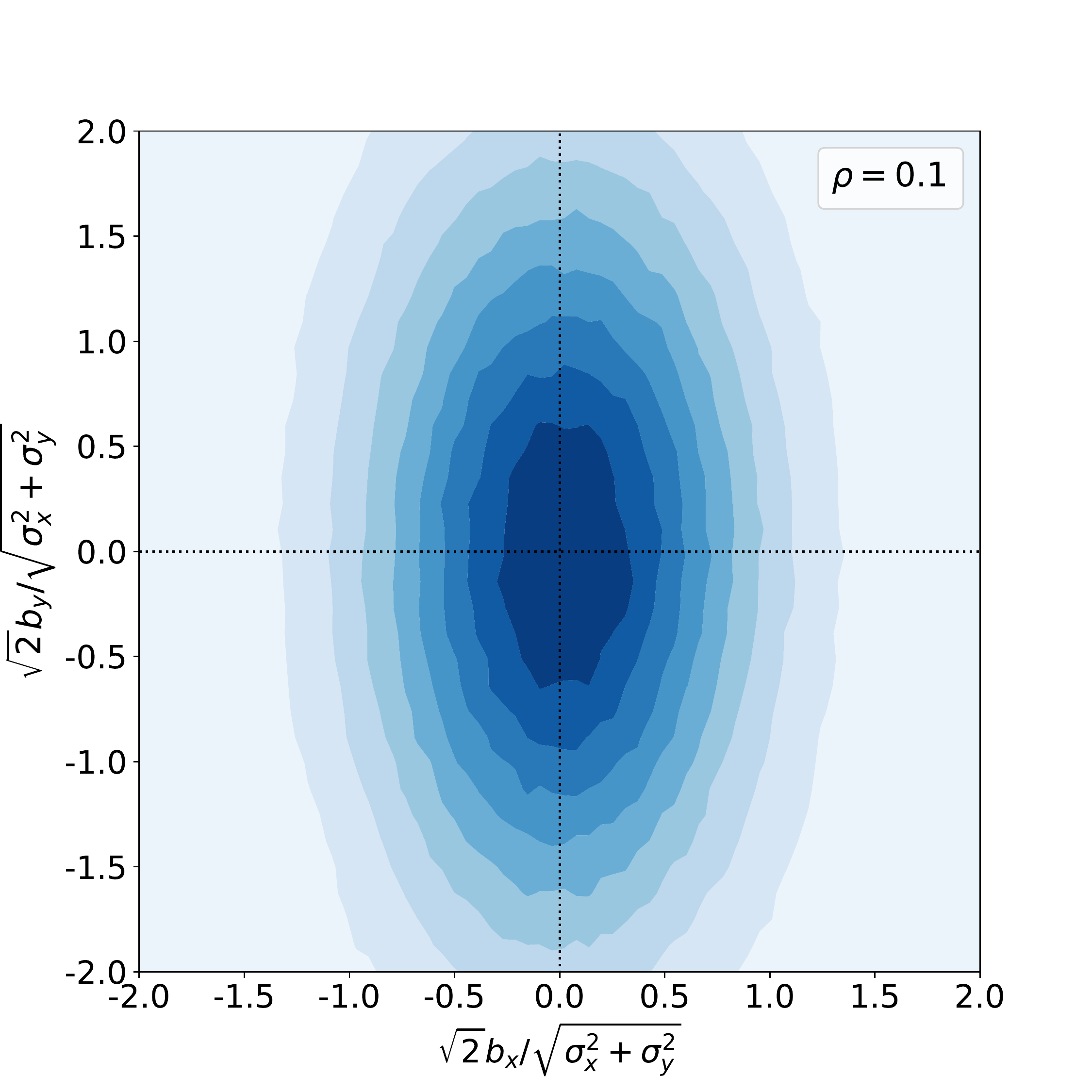}
    \caption{Global random GMF probability distribution.
    $\rho=1.0$ provides symmetric distribution between $b_x = \mathbf{b}\cdot\hat{\mathbf{x}}$ and $b_y = \mathbf{b}\cdot\hat{\mathbf{y}}$.
    $\rho=10$ corresponds to the parallel-aligned case where the $b_y$ is suppressed with respect to $b_x$.
    $\rho=0.1$ represents the perpendicular-aligned case where $b_x$ is suppressed with respect to $b_y$.
    $\sigma_{x,y}$ represents the root mean square (RMS) of $b_{x,y}$.}\label{fig:global_dist}
\end{figure}

Note that separating the divergence cleaning process from spatial re-profiling comes with a cost.
Strong alignment with $\rho \ll 1$ or $\rho \gg 1$ are not realizable because the Gram-Schmidt process reestablishes some extra spatial isotropy according to equation\,\eqref{eq:gs}.
Figure\,\ref{fig:global_dist} presents typical results of the global random generator in the form of magnetic field probability density distributions,
where we assume a Kolmogorov power spectrum.
The distributions of $b_y$ and $b_z$ are expected to be identical with the imposed alignment direction being $\hat{\mathbf{H}} = \hat{\mathbf{x}}$.
Note that the global generator is designed for realizing the inhomogeneity and anisotropy in both spatial and frequency domains,
which we then have to process with divergence cleaning to provide conceptually acceptable realizations.

\subsection{local random GMF generator}\label{sec:local_generator}

The local generator is proposed for realizing random GMFs in small scale regions, like the solar neighbourhood, where the regular field can be approximated as homogeneous with a uniform direction,
or more precisely speaking, where the random magnetic field 2-point correlation tensor can be approximated to be independent of the spatial position.
With this assumption, random fields can be realized with a single FFT.
Here we describe the vector decomposition method for realizing a Gaussian random magnetic field with a generic anisotropic power spectrum tensor $P_{ij}(\mathbf{k},\alpha)$,
where $\alpha$ represents extra parameters in addition to the wave-vector. 
By assuming Gaussianity the power spectrum tensor reads
\begin{eqnarray}
    P_{ij}(\mathbf{k},\alpha)\delta^3(\mathbf{k-k'}) &=& \langle \tilde{b}_i\mathbf{(k)}\tilde{b}_j^\ast\mathbf{(k')} \rangle_\mathbf{\tilde{b}} ~,\label{powerspec}
\end{eqnarray}
where $\tilde{\mathbf{b}}$ represents the complex magnetic field vectors in the frequency domain.
Depending on the specific form of the given power spectrum tensor, the vector field decomposition can be either explicit or implicit.

The implicit vector decomposition sets up two modes (vector bases) for a complex Fourier vector $\tilde{\mathbf{b}}$, which means
\begin{eqnarray}
    \tilde{\mathbf{b}}^{\pm}(\mathbf{k}) &=& \tilde{b}^{\pm}(\mathbf{k}) \hat{\mathbf{e}}^{\pm} ~,\\
    \hat{\mathbf{e}}^{\pm} &=& \frac{\hat{\mathbf{e}}_1\pm i\hat{\mathbf{e}}_2}{\sqrt{2}} ~,
\end{eqnarray}
where the two orthogonal basis vectors $\hat{\mathbf{e}}^{\pm}$ bind with the complex scalar $\tilde{b}^{\pm}$ respectively. The vectors $\{\hat{\mathbf{e}}_1,\hat{\mathbf{e}}_2,\hat{\mathbf{e}}_3\}$ form a Cartesian frame, and to ensure the divergence-free property of the resulting fields we choose $\hat{\mathbf{e}}_3 = \hat{\mathbf{k}}$.
During the Fourier transform of $\mathbf{\tilde{b}(k)}$ into the spatial domain we have to consider an orthogonal base aligned with the Cartesian grid of $\mathbf{b(x)}$, and here we adopt one convenient base representation as
\begin{eqnarray}
    \hat{\mathbf{k}} &=& (\frac{k_x}{k}, \frac{k_y}{k}, \frac{k_z}{k}) ~,\\
    \hat{\mathbf{e}}^- &=& (\frac{-k_y}{\sqrt{k_x^2+k_y^2}}, \frac{k_x}{\sqrt{k_x^2+k_y^2}},0) ~,\\
    \hat{\mathbf{e}}^+ &=& (\frac{k_xk_z}{k\sqrt{k_x^2+k_y^2}}, \frac{k_yk_z}{k\sqrt{k_x^2+k_y^2}}, \frac{-(k_x^2+k_y^2)}{k}) ~,
\end{eqnarray}
where $k = \sqrt{k_x^2+k_y^2+k_z^2}$.
Then we can proceed by projecting the complex field amplitude into this spatial frame
\begin{eqnarray}
    \tilde{\mathbf{b}}\cdot\hat{\mathbf{x}} &=& \tilde{b}^+(\hat{\mathbf{e}}^+\cdot\hat{\mathbf{x}}) + \tilde{b}^-(\hat{\mathbf{e}}^-\cdot\hat{\mathbf{x}}) ~,
\end{eqnarray}
where $\hat{\mathbf{x}}$ represents the spatial Cartesian coordinate.
Implicit decomposition is irrelevant to the choice of the $\{\mathbf{e}^+,\mathbf{e}^-\}$ base and useful in the case where only the spectrum trace $\mathrm{Tr}[P_{ij}(\mathbf{k})]$ (over the $i,j$ indices) is given.
The amplitude of $\tilde{b}^\pm$ can be drawn from Gaussian distributions with zero mean and variances $\sigma^2_{\pm}$ which satisfy
\begin{eqnarray}\label{eq:implicit}
    \sigma^2_{+} + \sigma^2_{-} &=& \mathrm{Tr}[P_{ij}(\mathbf{k})] d^3k ~,
\end{eqnarray}
with $d^3k$ represents the frequency domain discretization resolution.
equation\,\eqref{eq:implicit} indicates that the field amplitudes $\tilde{b}^{\pm}$ should have a joint power spectrum equal to the trace of the total power spectrum.

The explicit decomposition should be used when the power spectrum tensor is available along with the explicitly defined base $\{\mathbf{e}^+,\mathbf{e}^-\}$, where
\begin{eqnarray}
    \sigma^2_{\pm} &=& P^\pm(\mathbf{k}) d^3k ~.
\end{eqnarray}
A practical example is realizing Alfv{\'e}n, fast and slow modes of a MHD turbulent magnetic field in a compressible plasma.
Given a local regular GMF field $\mathbf{B}_0$,
an Alfv{\'e}n wave propagates along $\hat{\mathbf{B}}_0$ with magnetic turbulence in direction $\mathbf{e}^+=\hat{\mathbf{k}}\times\hat{\mathbf{B}}_0$ while slow and fast waves generate magnetic turbulence in direction $\mathbf{e}^-= \mathbf{e}^+\times \hat{\mathbf{k}}$.
A detailed parameterization of compressible MHD turbulent power spectrum will be introduced in Section\,\ref{sec:powerspec} following the corresponding references therein.
Note that when the wave-vector $\mathbf{k}$ is aligned with $\mathbf{B}_0$, the amplitudes of the Alfv{\'e}n and slow modes vanish and the fast mode realization requires an implicit decomposition as the base $\{\mathbf{e}^+,\mathbf{e}^-\}$ is undefined.

\begin{figure}
    \plotone{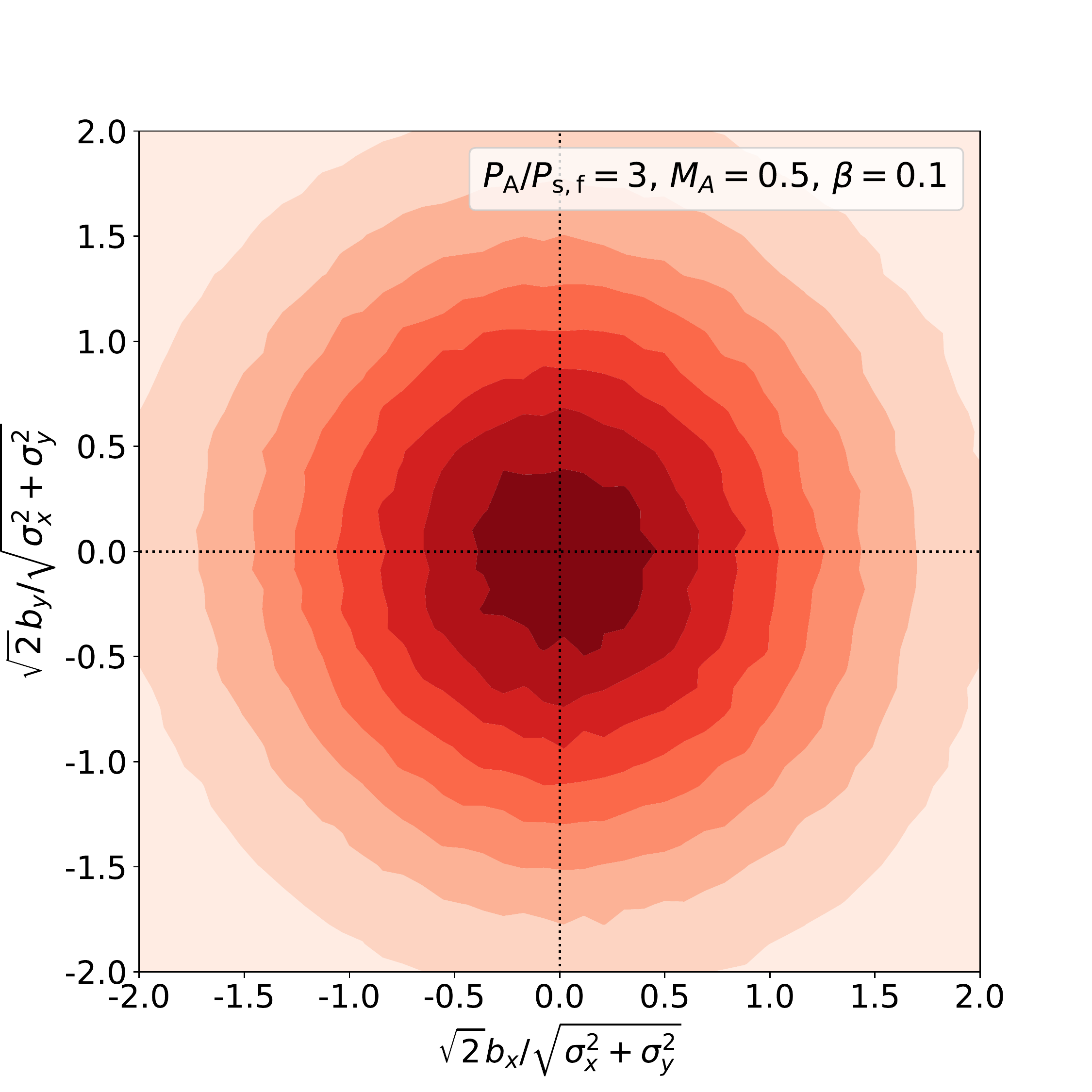}
    \plotone{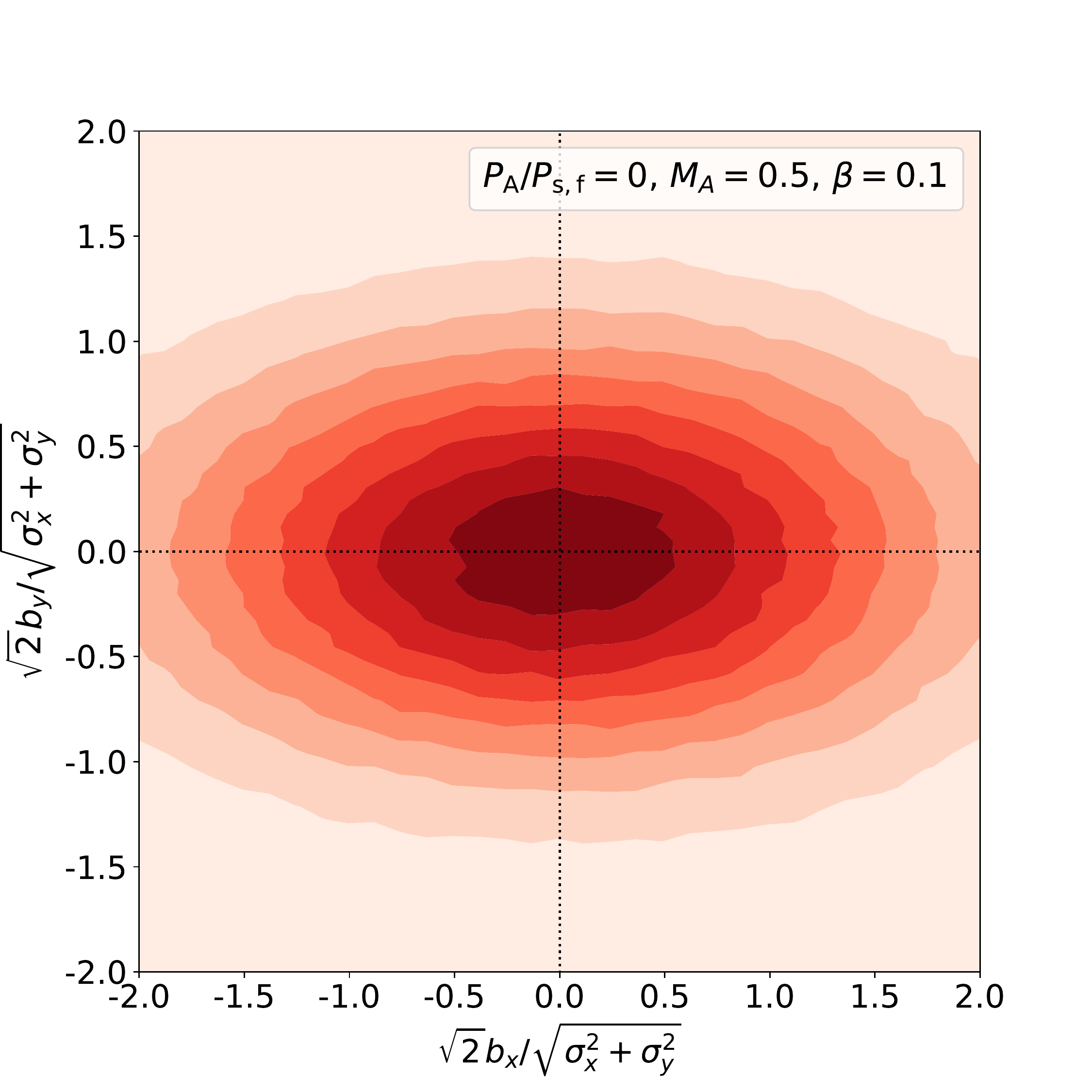}
    \plotone{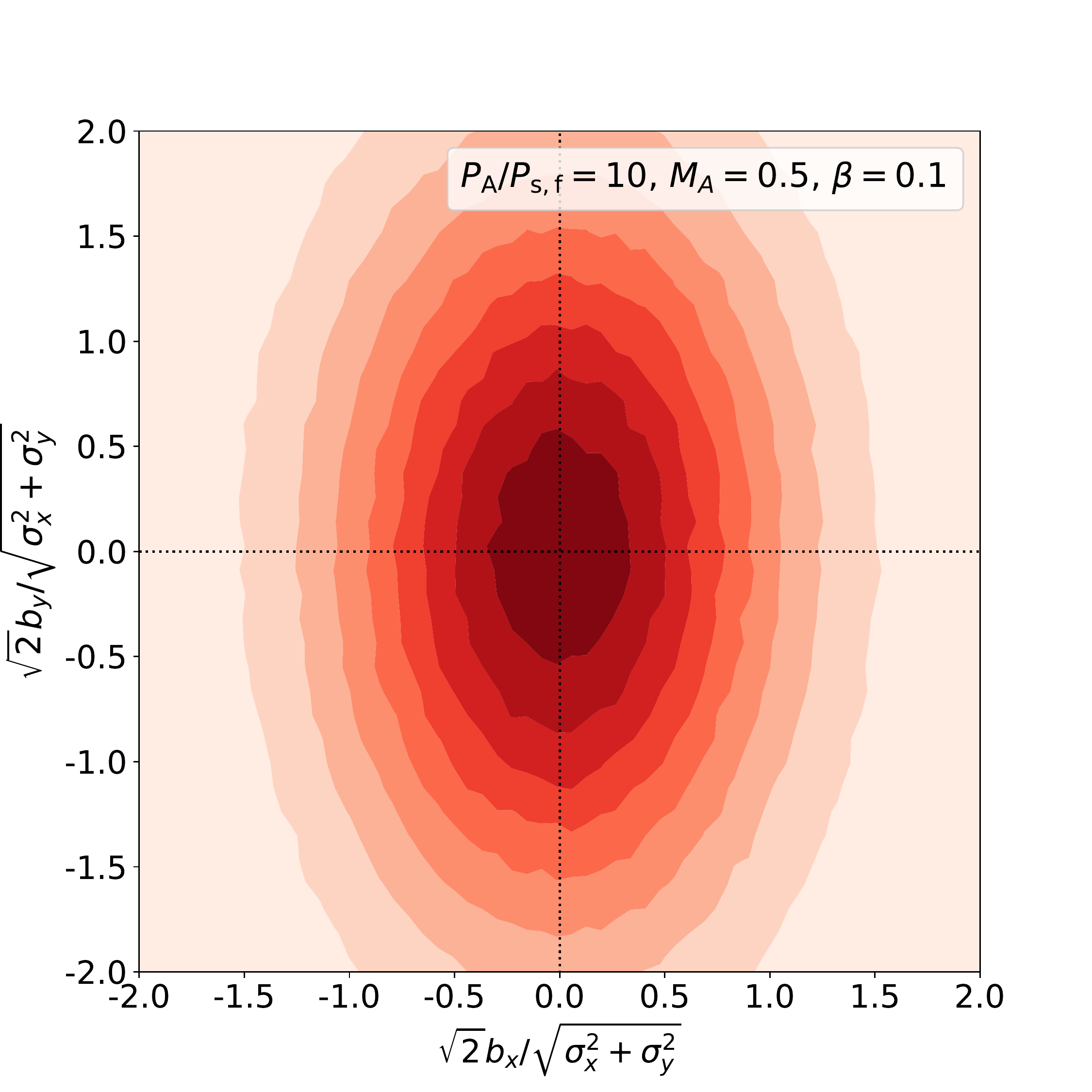}
    \caption{Local random field probability distribution with $\hat{\mathbf{B}}_0 = \hat{\mathbf{x}}$, Mach number $M_\mathrm{A}=0.5$, plasma parameter $\beta=0.1$. 
    $P_\mathrm{A}$ represents Alfv{\'e}n mode power at the injection scale, while for fast and slow modes we set equal power $P_\mathrm{f} = P_\mathrm{s}$ at the injection scale.
    $\sigma_{x,y}$ represents the RMS of $b_{x,y}$.}\label{fig:local_dist}
\end{figure}

Figure\,\ref{fig:local_dist} presents typical examples of the distribution of the random GMF from the local generator.
In comparison to the magnetic field distribution from the global generator where the spatial anisotropy is defined by the orientation alignment, the local generator is capable of realizing more subtle field properties, e.g., the spectral anisotropic MHD wave types described in Section\,\ref{sec:powerspec}.
At the phenomenological level, the global generator can mimic the random magnetic field orientation alignment of the local realizations as illustrated by Figure\,\ref{fig:global_dist} and Figure\,\ref{fig:local_dist}, but the spectral anisotropy is uniquely realizable by the local generator.

\newpage
\section{application example}\label{sec:powerspec}

To demonstrate the usefulness of \hammurabiX\ we investigate the properties of simulated synchrotron emission at high Galactic latitudes according to different random magnetic field configurations.
By focusing on the high latitude sky we concentrate on the properties of physical fields near the solar neighbourhood where both global and local random generators can be applied.

\citet{Alves2016} reported a synchrotron B/E ratio\footnote{The ratio between the B-mode and the E-mode of synchrotron angular power spectrum, i.e., $C^\mathrm{BB}_\ell/C^\mathrm{EE}_\ell$.} around $0.35$ at angular modes $l \in (30,300)$ (similar result has also been reported at high Galactic latitudes by \citet{Krachmalnicoff2018}),  which a successful modelling of the GMF should be able to explain.
Besides, a low polarization fraction at high Galactic latitudes is observed \citep{Ade2015}.
According to recent theoretical work by \citet{Kandel2018}, it may be possible to achieve a synchrotron B/E ratio lower than $1.0$ at high Galactic latitudes with compressible MHD turbulence, especially with slow and/or Alfv{\'e}n modes at low Mach number $M_\mathrm{A} < 0.5$.
An analytic calculation of the angular power spectrum observed in polarized synchrotron emission is not a trivial task.
As presented in theoretical estimations carried out by \citet{Caldwell2016}, \citet{Kandel2017} and \citet{Kandel2018}, it is impossible to avoid a certain level of simplification, e.g., the flat sky assumption, the Limber approximation, and the limitation of the perturbative regime.
Now with the help of \hammurabiX\ we can approach this topic numerically without being confined by the limits in analytic work.

To avoid distractions from other Galactic components or local structure models, in the following analyses, we assume a uniform distribution for the regular GMF parallel to the Galactic disk and a homogeneous CR electron density with a fixed spectral index.
No spatial modulation of the field strength is performed, but we use the ability to model the field orientation alignment described in Section\,\ref{sec:global_generator}.
The detailed modelling of MHD turbulence is briefly presented in the following.

\subsection{parameterized MHD turbulence}\label{sec:mhd}

A realistic formulation of the local turbulent GMF is essential in this work,
where simple random field generators usually cannot take into account the anisotropy imprinted on the wave-vector phases of the power spectrum.
The local generator we have designed in \hammurabiX\ is capable of carrying out a theoretical parameterization of MHD turbulent modes which have been discussed by \citet{Cho2002,Caldwell2016,Kandel2017,Kandel2018}.
As described in these references, the turbulent field power spectra for Alfv{\'e}n, fast and slow modes can be formulated as
\begin{eqnarray}
    P_i(k,\alpha) &=& P_i(k)F_i(M_\mathrm{A},\alpha)h_i(\beta,\alpha)~,\\
    P_i(k) &=& \frac{p_i}{4\pi k^2} \Big[
    \left(\frac{k_0}{k_1}\right)^{\alpha_1} \left(\frac{k}{k_1}\right)^6\mathcal{H}(k_1-k) \\\nonumber
    && + \left(\frac{k}{k_0}\right)^{-\alpha_1}\mathcal{H}(k-k_1)\mathcal{H}(k_0-k) \\\nonumber
    && + \left(\frac{k}{k_0}\right)^{-\delta_i}\mathcal{H}(k-k_0) \Big]~,\label{eq:delta}\\
    h_\mathrm{A} &=& 1~,\\
    h_\mathrm{f} &=& \frac{2}{D_{++}(1+\tan^2\alpha D^2_{-+}/D^2_{+-})}~,\\
    h_\mathrm{s} &=& \frac{2}{D_{-+}(1+\tan^2\alpha D^2_{++}/D^2_{--})}~,\\
    D_{\pm \pm} &=& 1\pm\sqrt{D}\pm 0.5\beta ~,\\
    D &=& (1+0.5\beta)^2-2\beta\cos^2 \alpha ~,\\
    F_\mathrm{f} &=& 1~,\\
    F_{\mathrm{A,s}} &=& \exp\left[-\frac{|\cos\alpha|}{(M^2_\mathrm{A}\sin \alpha)^{2/3}}\right]~,
\end{eqnarray}
where $i \in \{\mathrm{A},\mathrm{f},\mathrm{s}\}$ representing Alfv{\'e}n, fast and slow modes respectively\footnote{In this work the subscript $\mathrm{A}$ represents Alfv{\'e}n, $\mathrm{f}$ represents fast and $\mathrm{s}$ represents slow.}.
The two critical MHD parameters are the Alfv{\'e}n Mach number $M_\mathrm{A}$ and the plasma $\beta$ which is the ratio of gas pressure to magnetic pressure.
In the sub-Alfv{\'e}nic ($M_\mathrm{A}<1$) low-$\beta$ ($\beta<1$) regime, the spectral indices in equation\,\eqref{eq:delta} can be approximated as $\delta_\mathrm{A}=\delta_\mathrm{s}=5/3$, and $\delta_\mathrm{f}=3/2$ \citep{Cho2002}. 
The Alfv{\'e}n speed $v_\mathrm{A}$ which should appear in $h_i(\alpha)$ is absorbed by the normalization factor $p_i$ for simplicity.

\subsection{high latitude synchrotron emission}

\begin{figure}
    \includegraphics[width=0.35\textwidth]{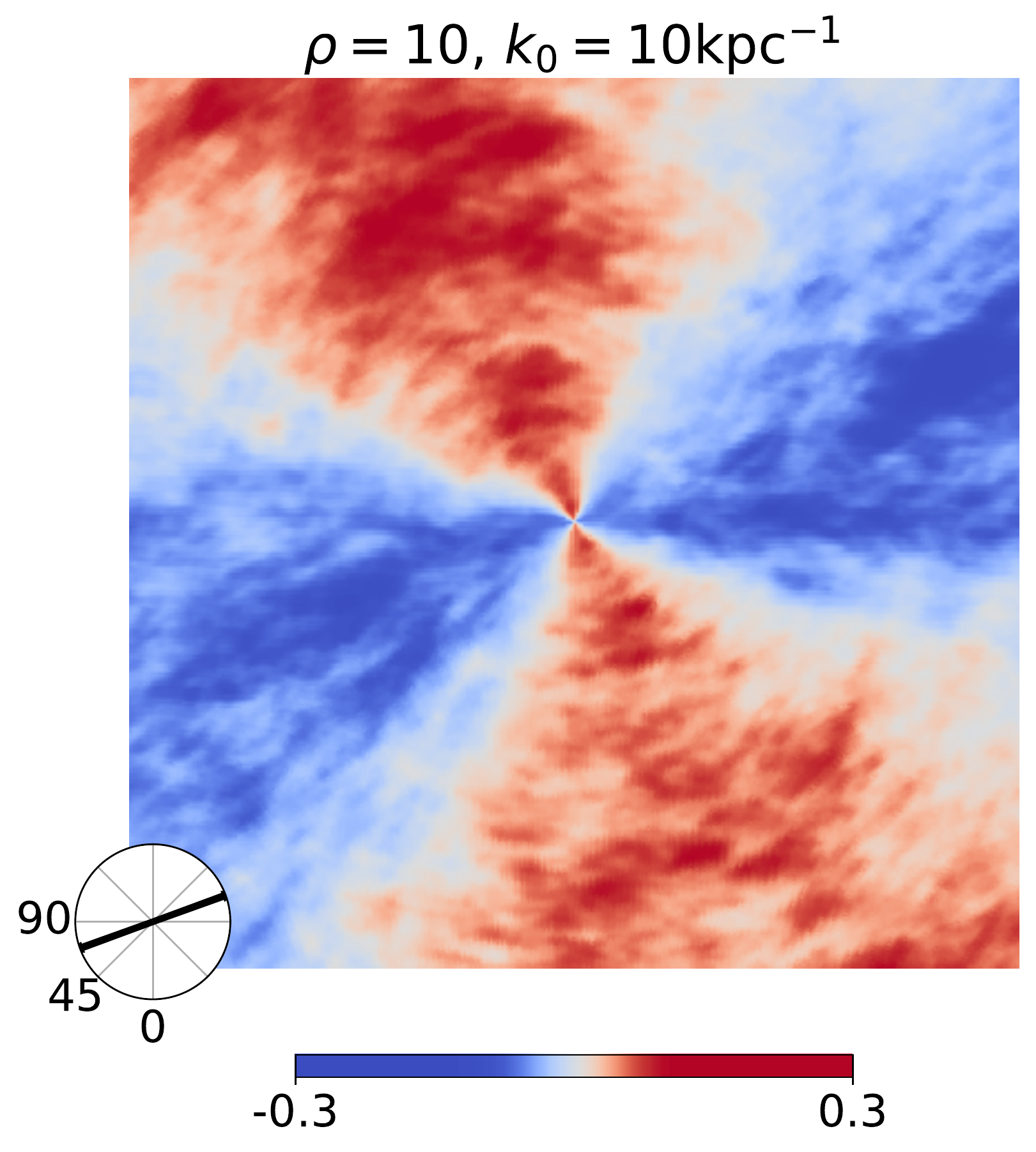}
    \includegraphics[width=0.35\textwidth]{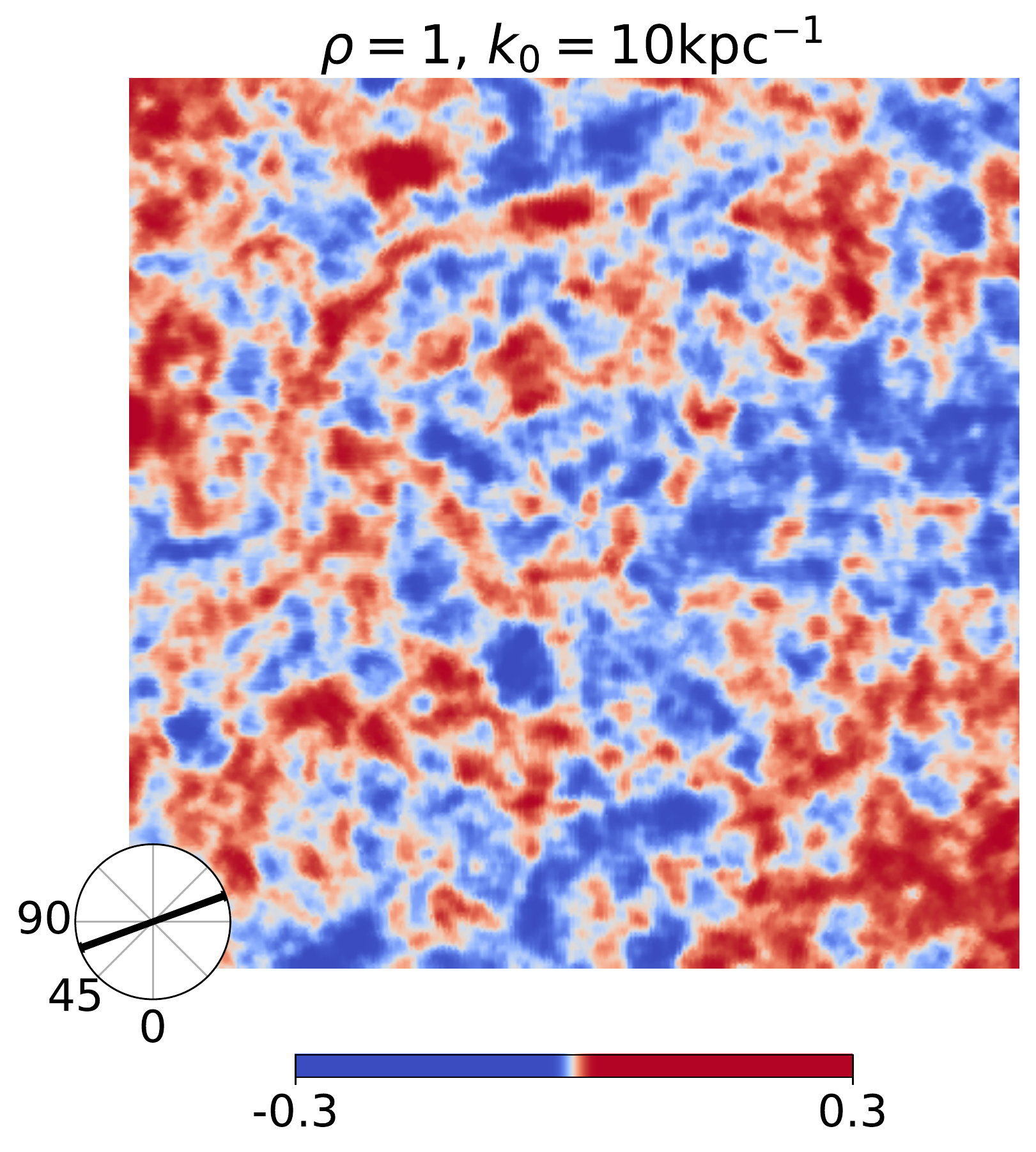}
    \includegraphics[width=0.35\textwidth]{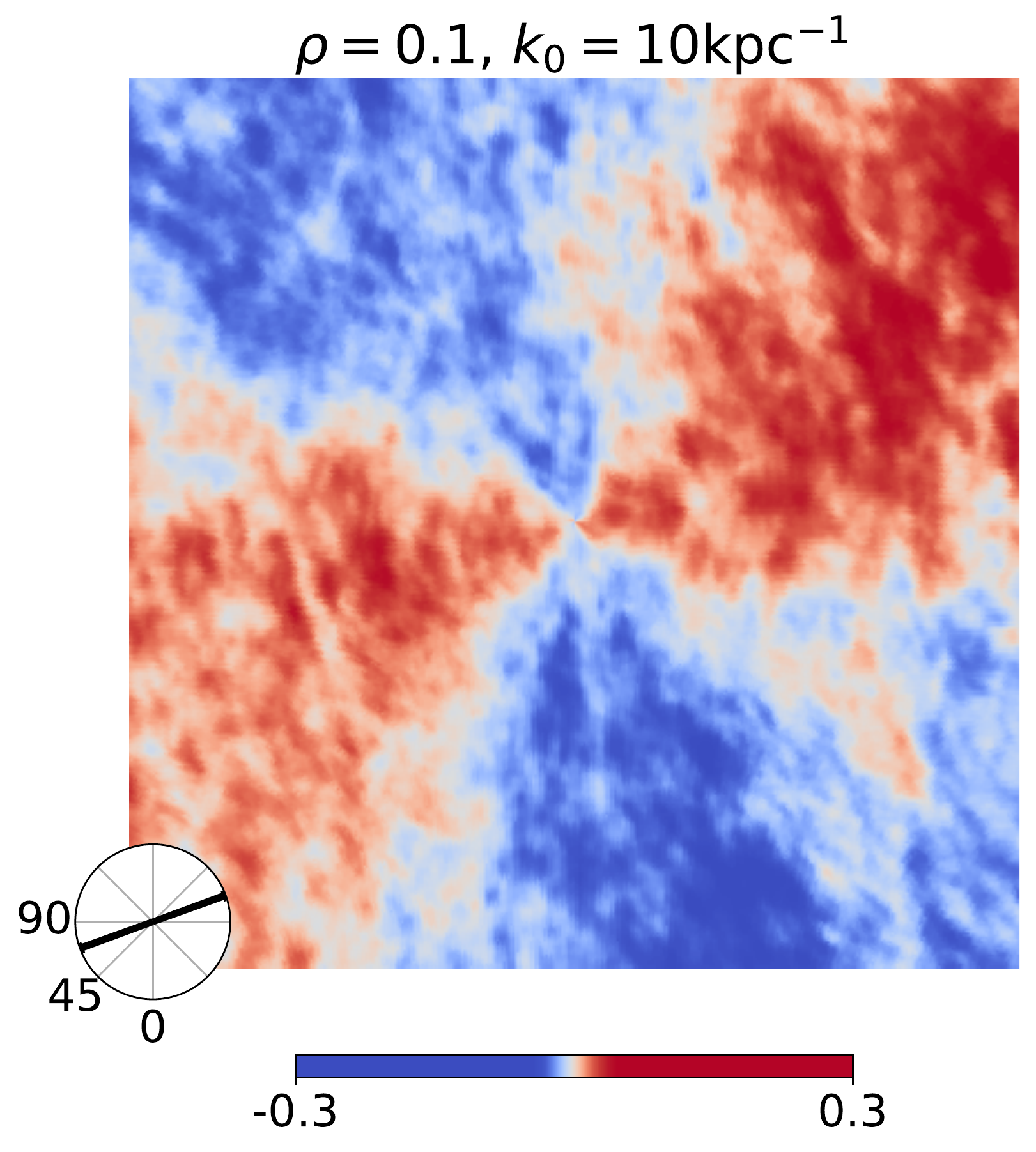}
    \caption{$30~\mathrm{GHz}$ synchrotron Stokes Q at the Galactic north pole in a 40 degree patch.
    The GMF simulation consists of a uniform regular (with orientation displayed on the bottom-left corner of each panel) and global random component with injection scale $k_0=10~\mathrm{kpc}^{-1}$ but different alignment parameter $\rho=10$ (top), $\rho=1$ (middle) and $\rho=0.1$ (bottom).
    The strength ratio between the random and regular GMF is $b/B_0=3.0$.}\label{fig:unif_gr_k10_q}
\end{figure}

\begin{figure}
    \includegraphics[width=0.35\textwidth]{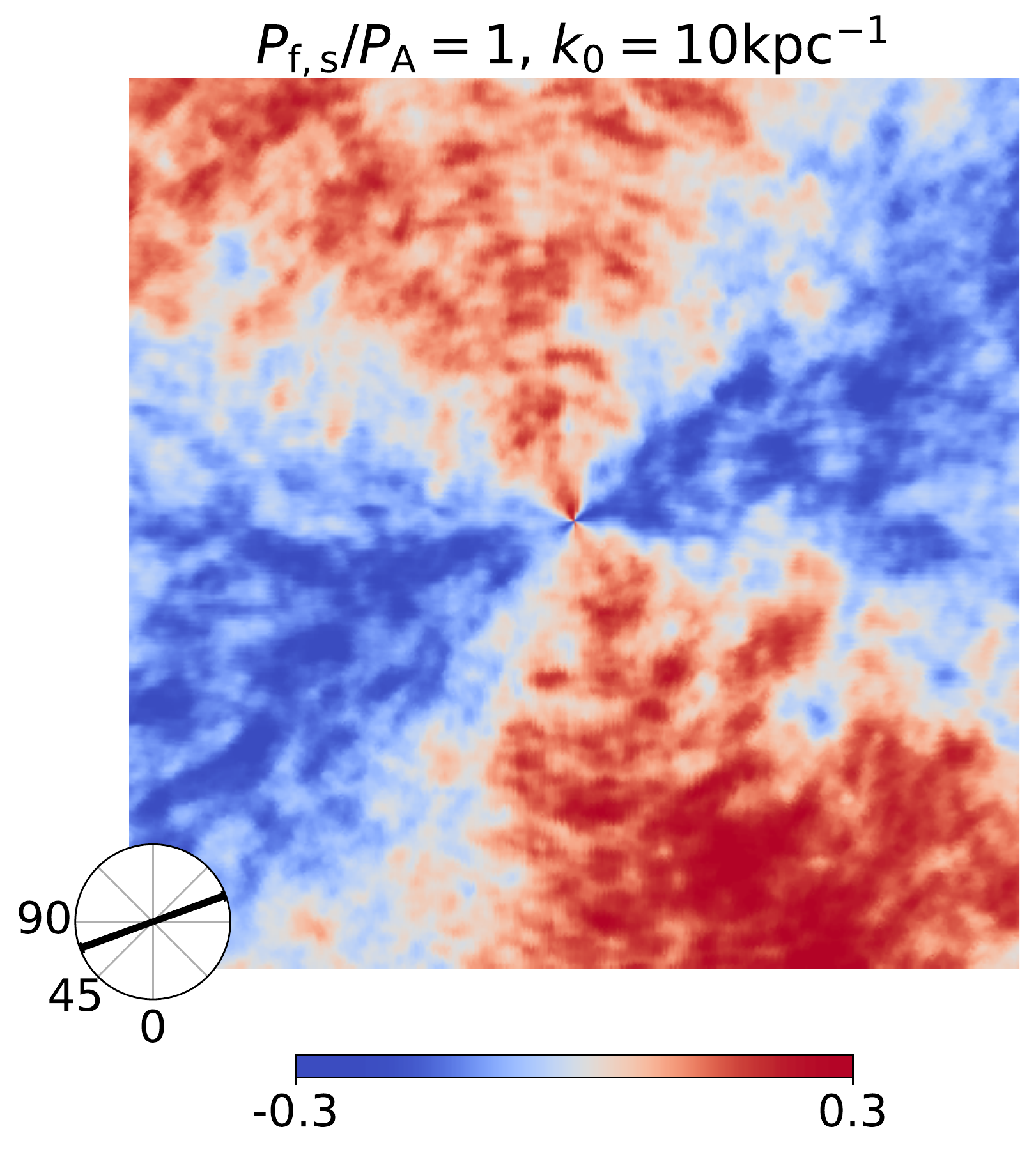}
    \includegraphics[width=0.35\textwidth]{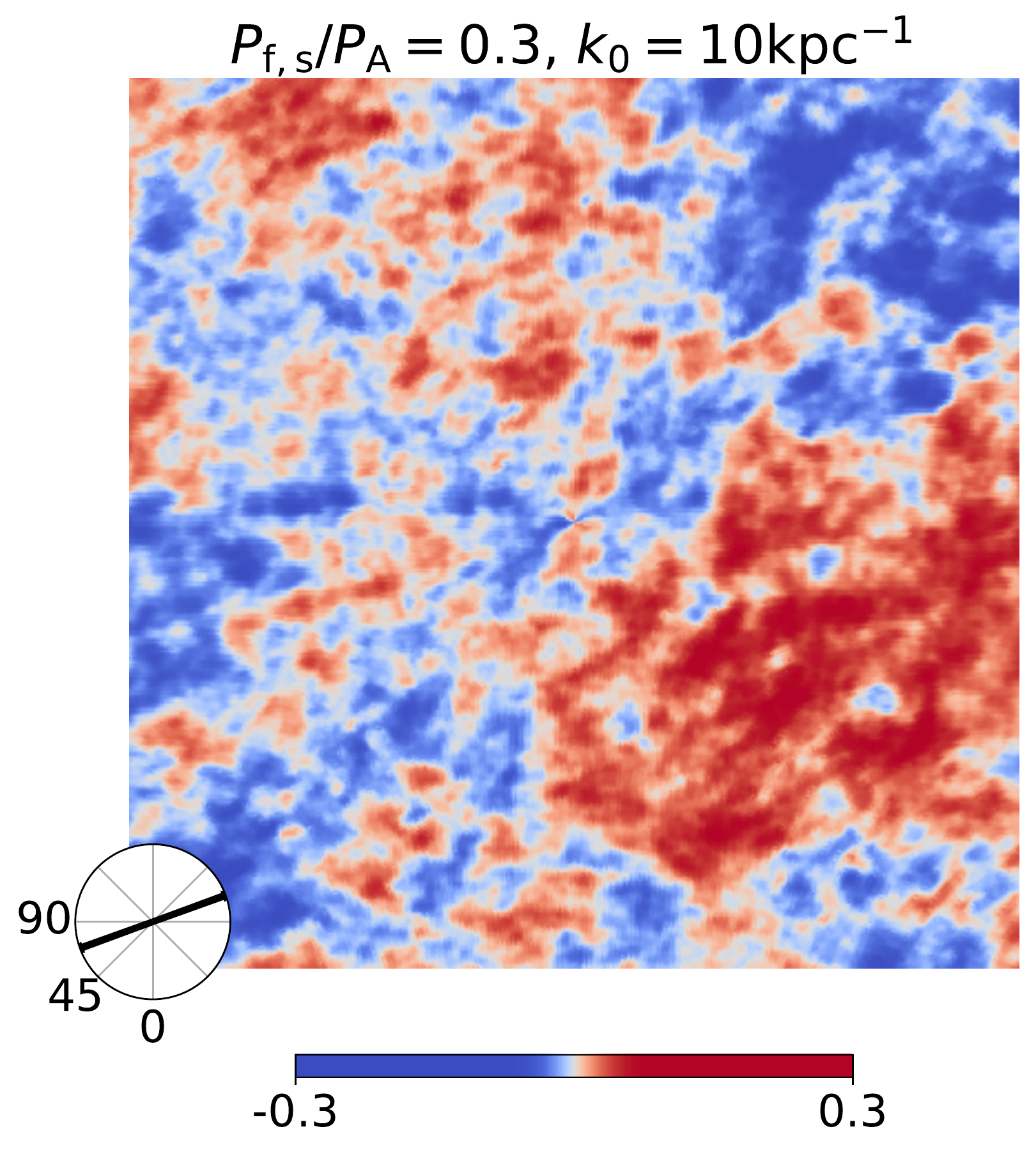}
    \includegraphics[width=0.35\textwidth]{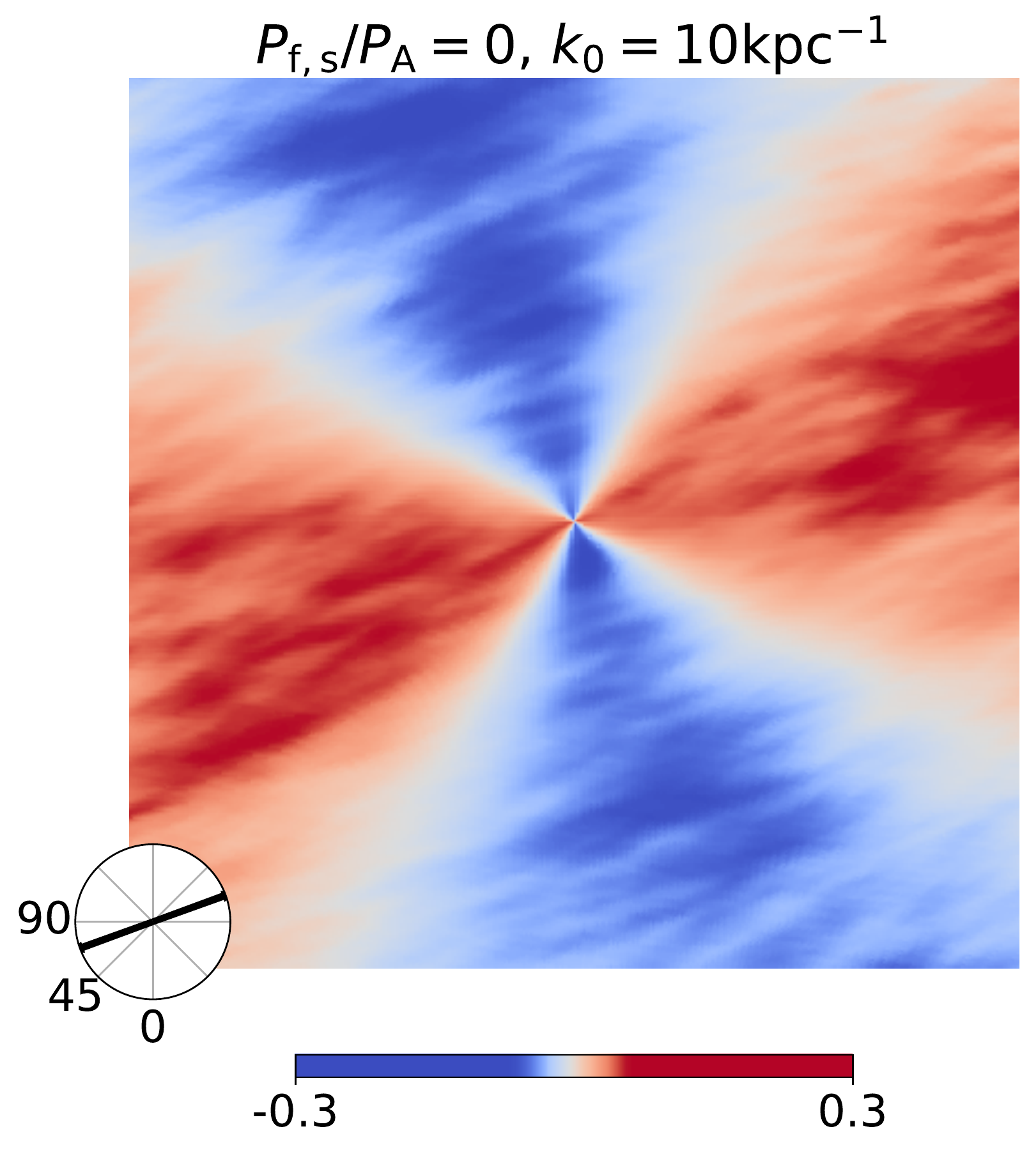}
    \caption{$30~\mathrm{GHz}$ synchrotron Stokes Q at the Galactic north pole in a 40 degree patch.
    The GMF simulation consists of a uniform regular (with orientation displayed on the bottom-left corner of each panel) and a local random component with total spectral power $k_0P_0/B^2_0=75.0$ at injection scale $k_0=10~\mathrm{kpc}^{-1}$.
    The Alfv{\'e}n Mach number $M_\mathrm{A}=0.5$ and plasma parameter $\beta=0.1$ are set to match the parameterization in Figure\,\ref{fig:local_dist}.}\label{fig:unif_lr_k10_q}
\end{figure}

With the improved precision in \hammurabiX, we present high-resolution Galactic synchrotron emission simulations with analytic models as described above.
Presented in Figure\,\ref{fig:unif_gr_k10_q} are the examples of synchrotron polarization at high Galactic latitudes predicted by a uniform regular GMF parallel to the Galactic disk and a random component from the global generator with a Kolmogorov power spectrum.
Maps of synchrotron polarization from the same regular GMF but the local generator using a compressible MHD model are presented in Figure\,\ref{fig:unif_lr_k10_q}.
Since we are presenting only illustrative models, the absolute strength of regular and random GMF is not essential here.

The most prominent feature of the high latitude synchrotron polarization is the quadrupolar structure that results from the GMF orientation at the solar neighbourhood.
As the examples displayed in Figure\,\ref{fig:unif_gr_k10_q}, the quadrupole direction is largely determined by the regular field, but on top of which we can observe a flip in the polarization between the regimes when $\rho>1.0$ versus $\rho<1.0$.
When the random GMF has no preferred alignment, i.e., the $\rho=1.0$ case, the quadrupole pattern is undermined by the isotropic random field contribution.
This is visually clear because the random field strength dominates.
In Figure\,\ref{fig:unif_lr_k10_q} the quadruple pattern is well preserved with MHD turbulence injection scale $k_0 = 10~\mathrm{kpc}^{-1}$, and also a flip in the polarization can be observed with the pure Alfv{\'e}n mode when the random field dominates.
When the spatial distribution or random GMF is close to spatially isotropic\footnote{The local generator has no field alignment parameter like $\rho=1.0$ that can ensure an absolutely spatially isotropic distribution with respect to $\mathbf{B}_0$.} with $P_\mathrm{A}/P_\mathrm{f,s}=3.0$ (and Alfv{\'e}n Mach number $M_\mathrm{A}=0.5$, plasma parameter $\beta=0.1$) as displayed by the top panel in Figure\,\ref{fig:local_dist},
we observe a similar trend of weakening the quadrupole pattern as demonstrated by Figure\,\ref{fig:unif_lr_k10_q}.

\begin{figure}
    \plotone{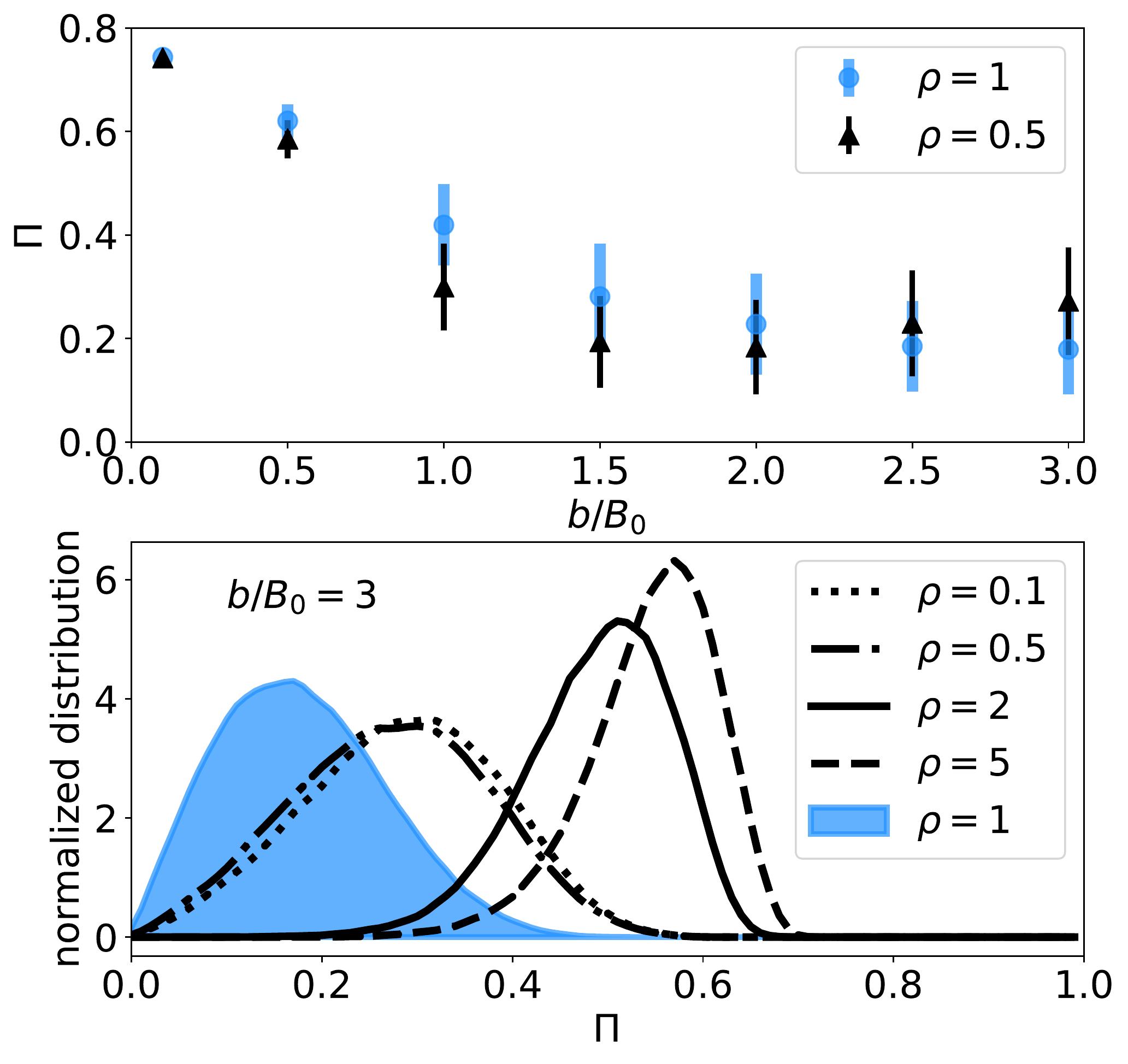}
    \caption{Distribution of synchrotron polarization fraction $\Pi$ at high Galactic latitudes produced by a uniform regular and global random GMF.
    In the top panel, the distribution (16th to 68th percentile) characterized by mean and standard deviation as a function of random field strength is displayed, where the alignment ratio is fixed.
    In the bottom panel, we show a histogram of polarization fraction where $b/B_0=3.0$ and the alignment parameter $\rho$ varies. Recall that $\rho=1$ is isotropic while $\rho<1$ and $\rho>1$ are anisotropic.}\label{fig:global_polr}
\end{figure}

\begin{figure}
    \plotone{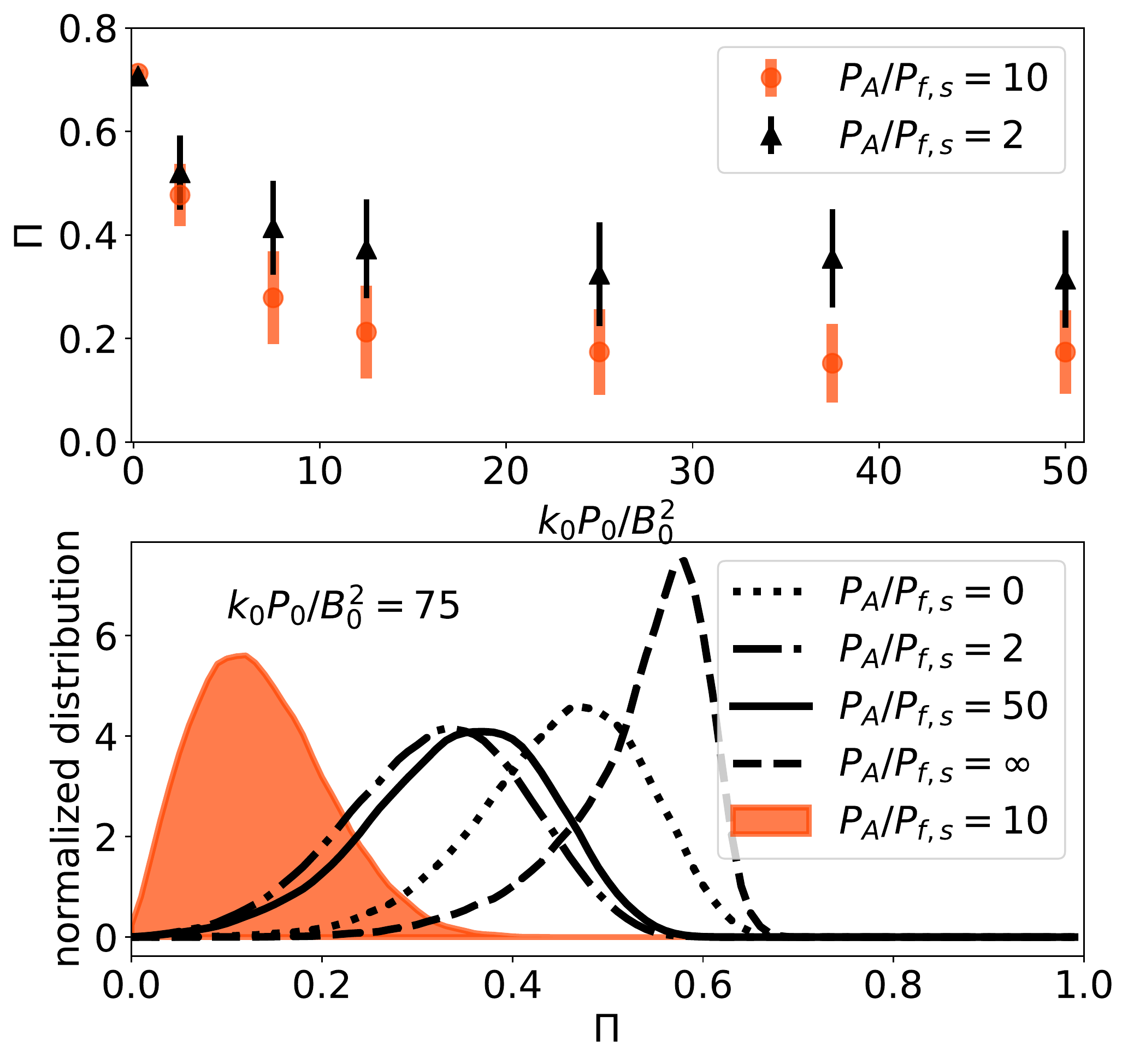}
    \caption{Distribution of synchrotron polarization fraction $\Pi$ at high Galactic latitudes produced by uniform regular and local random GMF.
    In the top panel, the distribution (16th to 68th percentile) characterized by mean and standard deviation as a function of random field strength is displayed, where the anisotropy ratio $P_\mathrm{A}/P_\mathrm{f,s}$ is fixed at the injection scale $k_0=10~\mathrm{kpc}^{-1}$ while the ratio between the total spectral power $P_0=P_\mathrm{f}+P_\mathrm{s}+P_\mathrm{A}$ at the injection scale and the regular field energy $P_0/B^2_0$ varies.
    In the bottom panel, $k_0P_0/B^2_0=75.0$ while the anisotropy ratio $P_\mathrm{A}/P_\mathrm{f,s}$ varies.} \label{fig:local_polr}
\end{figure}

The synchrotron polarization fraction (or the degree of linear polarization) is mainly determined by the CRE spectral shape when a uniformly distributed regular GMF dominates.
Assuming a constant CRE spectral index $\alpha=3.0$, the synchrotron polarization fraction $\Pi = (3\alpha + 3)/(3\alpha + 7)$ is much higher than that observed from Planck data \citep{Adam2016}.
Figures\,\ref{fig:global_polr} and \ref{fig:local_polr} demonstrate that the synchrotron polarization fraction can be suppressed by a Gaussian random field as long as the random field is not strongly anisotropic in the spatial domain.
The suppression in polarization fraction grows with the increasing of random field strength but depends on the specific field modelling.
Recall that the addition of a random component to the magnetic field direction functions as a random walk in the polarization plane, which means that even for a purely turbulent field, the polarized intensity continues to increase with the number of turbulent cells added along the LoS.
In principle, the increase goes as the square-root of the number of cells, while the total intensity increases linearly, so the fraction should decrease accordingly.
In practice, the precise trend is complicated by the effect of the observational beam and the locally varying anisotropy.
The shape of the polarization fraction for the $\rho=0.5$ model in Figure\,\ref{fig:global_polr}, for example, is due to the anisotropic random field cancelling with the regular field before beginning to dominate.
An inhomogeneous distribution (by field strength modulation) of the random field can change the efficiency of suppression differently depending on the field alignment, but the common features described above are preserved.

The above analyses imply that interpreting the synchrotron polarization toward the poles as due to the local field direction neglects the possible effects of anisotropic turbulence, which can mimic or flip the morphology.
Though the physical process is different, the geometry of the field and its effect on the observables is the same for polarized dust emission.
This work illustrates the opportunity for retrieving useful information of local magnetic turbulence structure with high latitude Galactic polarized emission, and also shows the challenge from the degeneracy between random and regular magnetic field orientations when using emission data alone.
It suggests that we need to be careful about realizing the local GMF structure to avoid misleading conclusions.
For example, it has been proposed recently by \citet{Alves2018} that according to observations, the regular magnetic field structure may play a dominant role in Galactic dust emission near the solar neighbourhood.
We also emphasize that the Galactic synchrotron emission is also affected by the warm ISM in the Galactic thick disk and even the halo.
The random field generators in \hammurabiX\ can be used to bridge the gap between simple large-scale field models and computationally intensive MHD simulations, and push toward more realistic analysis and modelling than previous methods.

\subsection{angular power spectrum}\label{sec:Cls}

The large angular scale Galactic synchrotron polarization pattern driven mainly by the GMF orientation at the solar neighbourhood is quite evident as illustrated in Figures \ref{fig:unif_gr_k10_q} and \ref{fig:unif_lr_k10_q}.
However, the small angular structures can be analyzed with the angular power spectrum, which can be decomposed by rotation-invariant components, i.e., the T, E and B modes \citep{Hu1997p}.
With the two random field generators proposed in this work, we intend to figure out which properties of the random GMF are imprinted on the synchrotron B/E ratio.
Specifically, we are interested in verifying whether MHD turbulence modes are capable of producing $\mathrm{B}/\mathrm{E} < 1.0$ in both the perturbative and the non-perturbative regimes.
Since we are focusing on high latitude polarization, pixels at Galactic latitude within $\pm 60^\circ$ are masked out.
We also set a lower limit to the radius in the LoS integral according to the random field grid resolution and the spherical mode range.
Technical details of the precision checks for the pseudo-$C_\ell$ estimation is discussed in Appendix\,\ref{sec:cl_tech}.

We present in Figure\,\ref{fig:unif_gr_becurve} the B/E ratio distribution (by collecting results from an ensemble of realizations with each given parameter set) for varying random field strengths and alignments of the global random GMF.
Figure\,\ref{fig:unif_gr_becurve} implies that to reproduce $\mathrm{B}/\mathrm{E} < 1.0$ we either need random GMF in the non-perturbative regime ($b/B_0 > 1.0$) or parallel alignment ($\rho>1.0$).
We also note that the divergence cleaning step is what leads to $\mathrm{B}/\mathrm{E} \neq 1.0$.
As illustrated in the same figure, all realizations end up with $\mathrm{B}/\mathrm{E} = 1.0$ regardless of random field alignment, when the Gram-Schmidt process is switched off. 
This is expected given that a simple Gaussian random field should have $E = B$ on average, whereas a magnetic field must be divergence-free and therefore the difference between the naive random vector field and the magnetic field, which has been ignored in many previous analyses, is crucial in studying Galactic emissions.
Now we conclude that the divergence-free random magnetic field can provide synchrotron $\mathrm{B}/\mathrm{E} \ne 1.0$.
The Gram-Schmidt cleaning method is computationally useful and correct for reproducing the divergence-free random magnetic field (which in the simplest case can alternatively be obtained from a Gaussian random vector potential as shown in Appendix\,\ref{sec:alternative_tech} where synchrotron $B/E < 1$ arises naturally out of either method in the non-perturbative regime) and has the added benefit that we can spatially modulate its strength and orientation.

\begin{figure}[htb!]
    \plotone{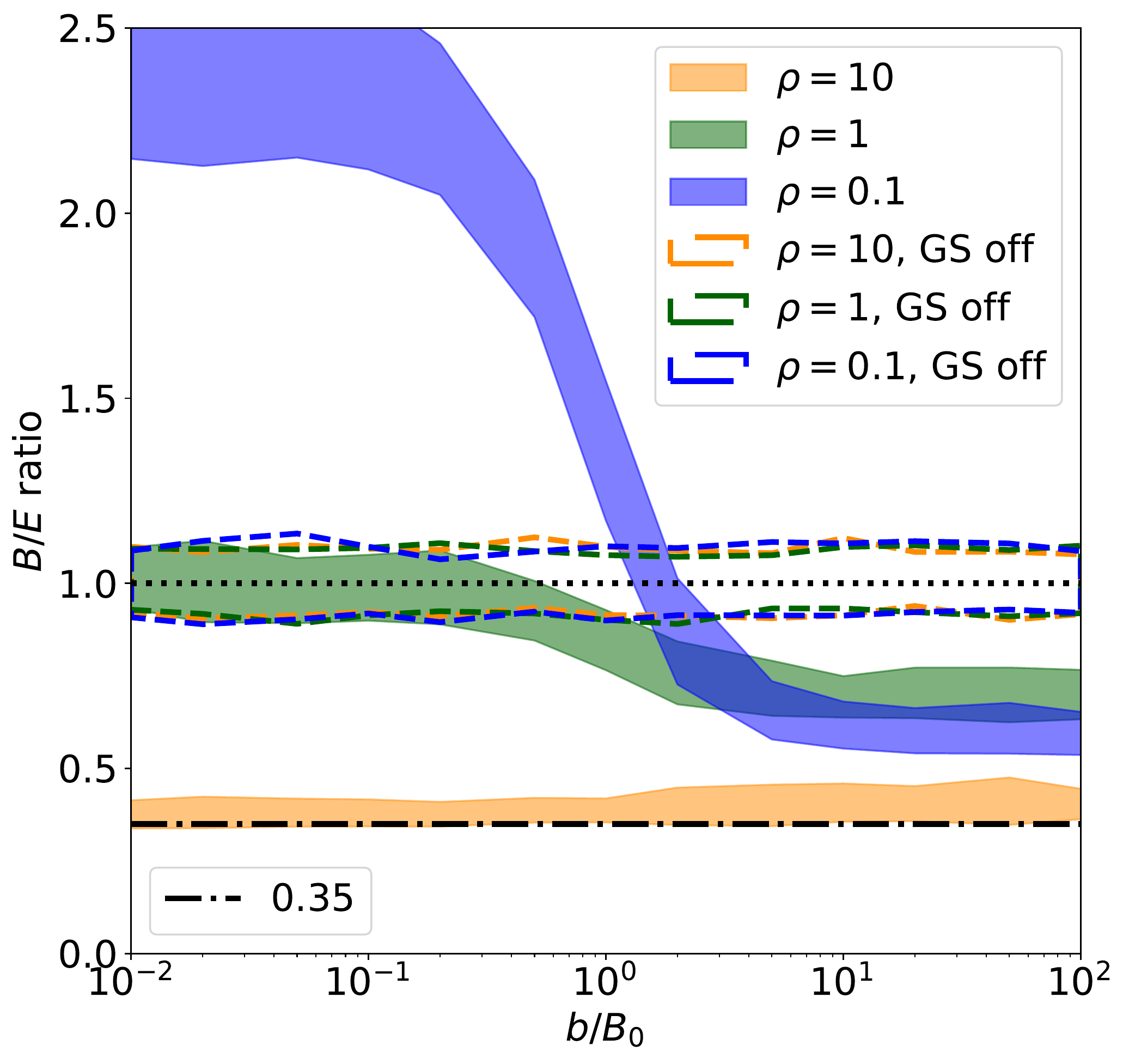}
    \caption{Distribution (16th to 68th percentile) of the $30~\rm{GHz}$ synchrotron emission B/E ratio for $\ell>100$ according to global random GMF with various field strength and alignment.
    The ensemble size is set as ten independent realizations at each sampling position, beyond which we found no significant improvement in the B/E estimation.
    The results marked by ``GS off'' come from random GMF without divergence cleaning.
    The contribution to the angular power spectrum from the regular GMF has been subtracted, which would otherwise dominate the B/E ratio in the perturbative regime ($b \ll B_0$).}\label{fig:unif_gr_becurve}
\end{figure}

\begin{figure}[htb!]
    \plotone{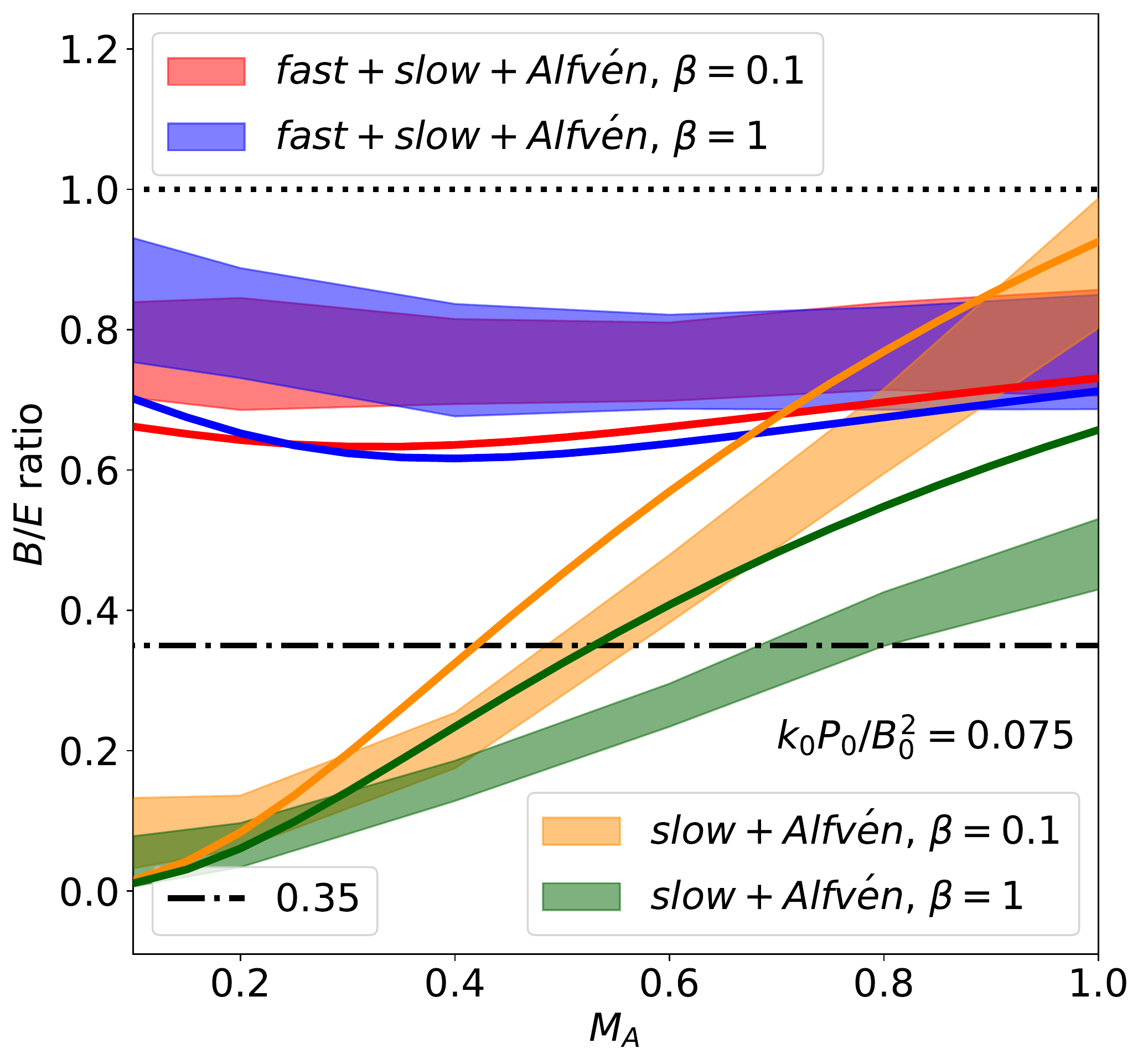}
    \plotone{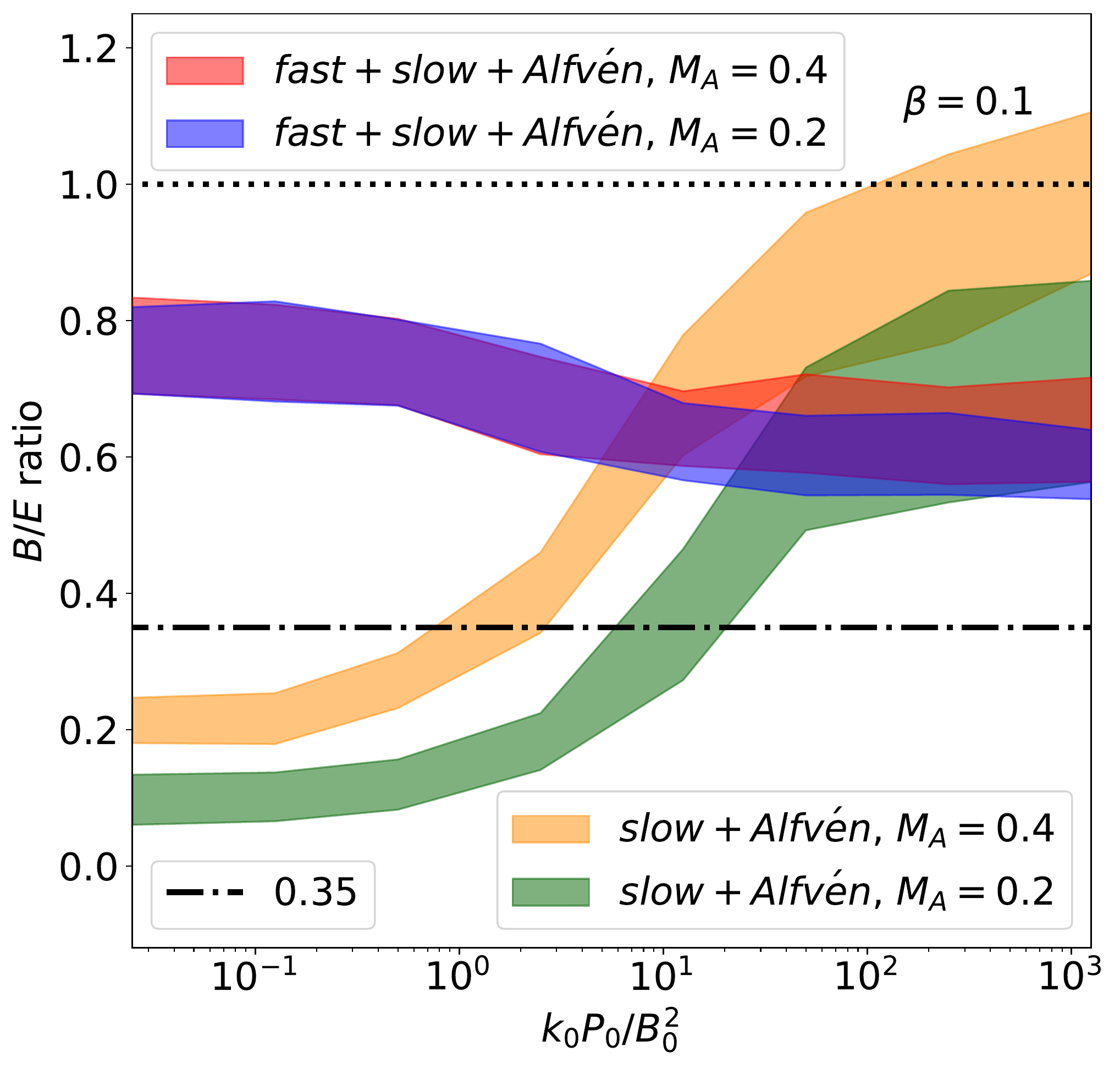}
    \caption{Distribution (16th to 68th percentile) of the $30~\rm{GHz}$ synchrotron emission B/E ratio for $\ell>100$ according to the local GMF realizations with various field strengths, Alfv{\'e}n Mach numbers, and plasma parameters.
    The ensemble size is set as ten independent realizations at each sampling position, beyond which we found no significant improvement in the B/E estimation.
    Solid lines in the top panel are predictions from \citet{Kandel2018}.
    The fast+slow+Alfv{\'e}n case sets equal magnetic field power at the injection scale for the three modes (i.e., $P_\mathrm{A}/P_\mathrm{f,s} = 1.0$), while the fast mode is excluded from the slow+Alfv{\'e}n case (i.e., $P_\mathrm{s} = P_\mathrm{A}$).
    The contribution to the angular power spectrum from the regular GMF has been subtracted, which would otherwise dominate the B/E ratio in the perturbative regime ($k_0P_0 \ll B^2_0$).}\label{fig:unif_lr_becurve}
\end{figure}

By contrast, the $C_\ell$s estimated from the local MHD realizations have a clear analytic representation, to which simulations can be directly compared.
To look for the low B/E ratio according to \citet{Kandel2018}, we keep the random GMF strength at the perturbative level and tune the MHD Mach number $M_\mathrm{A}=0.2$ and plasma parameter $\beta=0.1$.
As illustrated in Figure\,\ref{fig:unif_lr_becurve}, we find clear evidence that a Gaussian realization of MHD turbulence can provide a synchrotron B/E ratio smaller than $1.0$, in both perturbative and non-perturbative regimes.
The fast mode in a sub-Alfv{\'e}nic low-$\beta$ plasma has a unique power spectrum shape and is less affected by the anisotropy function $h(\alpha)$ than the slow mode.
By assuming equal power in the turbulence modes at the injection scale, the observed angular power spectra are mainly influenced by the fast mode and so the B/E ratio has different behaviour for the case where slow and Alfv{\'e}n modes dominate.
With the given MHD Mach number and plasma parameter, slow mode turbulence results in a much lower B/E ratio than that from the Alfv{\'e}n mode, while fast mode prefers $\mathrm{B}/\mathrm{E} \simeq 0.8$ in perturbative regime.
These features are conceptually consistent with analytic predictions by \citet{Kandel2018} as demonstrated in the top panel of Figure\,\ref{fig:unif_lr_becurve}, where the differences between two estimations are likely because of the simplification in analytic derivation, e.g., the Limber and flat-sky approximations.
Beyond the perturbative regime, we observe the B/E ratio evolves with the growth of random field strength and suggests an upper limit for the random field strength to achieve the observed B/E ratio with solely MHD turbulence.

The observational implications of the Galactic synchrotron emission from above two types of random field realizations are that both the divergence-free and MHD turbulent nature of the field are important for producing synchrotron $\mathrm{B}/\mathrm{E}<1.0$ (aside from the fact that the divergence-free condition is physically required).
It is possible to use directly the angular power spectra estimated in the way presented here for studying Galactic components like the work by \citet{Vansyngel2018}, but we should be aware of the numeric uncertainty if the simulation resolution is lower than that of astrophysical measurements, in addition to the fundamental difference between simulation and observation mentioned in Section\,\ref{sec:hamx}.

\section{summary}\label{sec:sum}

In this report, we have presented \hammurabiX, the improved version of \hammurabi.
We have redesigned the package properly with calibrated precision and multi-threading support.
This report focuses on the implementation of the synchrotron emission simulation in \hammurabiX\ and its relation to the random magnetic field realization.
The technical features and profiles associated with Galactic synchrotron emission have been, for the first time, reported in detail.  

Two fast methods for generating divergence-free Gaussian random magnetic fields covering either Galactic scales or a local region have been proposed.
This is a crucial improvement (in computing accuracy and the capability of realizing physical features) over not only the previous versions of \hammurabi\ but also previous fast methods of simulating the GMF and the resulting diffuse Galactic polarized emission from the ISM.
It is increasingly clear that simplistic treatments of the turbulent component of the ISM do not produce simulated observables of sufficient complexity to be useful in comparison to the data.
Though full MHD turbulence realizations are computationally too expensive for the usage in large-scale GMF model fitting, using the statistical properties of these MHD simulations is an important intermediate step pursued here.
The new \hammurabiX\ provides the ability for the first time to generate Gaussian simulations that capture some of the properties of fast, slow, and Alv\'en modes of MHD turbulence in a computationally efficient approximation.
Using these more realistic numerical methods for simulating the magnetized ISM will lead to results that can be more directly linked to physical theory.

We have further demonstrated the importance of these improvements by studying two properties of the GMF that have been discussed in the literature.
Firstly, we have shown the importance of including a treatment of the anisotropic turbulence in the local ISM when attempting to interpret high-latitude synchrotron polarization as an indication of the local magnetic field direction.
Any such modelling of the local field can use \hammurabiX\ to quantify how much this affects the results, particularly with the addition of Faraday depth to break the degeneracy of using only polarized diffuse emission.
Secondly, using our new numerical methods, we have found that a Gaussian random realization with either the global field orientation alignment or the local MHD parameterization can produce $\mathrm{B}/\mathrm{E} \simeq 0.35$ in synchrotron emission at high Galactic latitudes.
Comparing the B/E ratio predicted by the global random GMF realizations with and without invoking the Gram-Schmidt process, we have realized that the divergence-free property is essential for such detailed statistical studies of GMFs.
Our results conceptually confirm the prediction made by \citet{Kandel2018} for Galactic synchrotron emission, which says the MHD magnetic turbulence has the ability to predict $\mathrm{B}/\mathrm{E} < 1.0$, while the prediction for dust emission B/E ratio has been conceptually confirmed by \citet{Kritsuk2018}.
We have also succeeded in demonstrating the computing power that \hammurabiX\ can provide to go beyond analytic studies of Galactic foreground observables with non-perturbative random GMF realizations.

In the near future, we would like to focus on improving the random GMF generators with more physical features.
The alignment of the random GMF around local filaments (including helicity) and non-Gaussianity will be interesting extensions, through which we can study the joint effect of the magnetic field alignment and its spectral anisotropy.
In \hammurabiX, both the global and local generators are designed to allow in the future the addition of non-Gaussianity, e.g., with the method introduced by \citet{Vio2001}, helicity, e.g., with the method instructed by \citet{Kitaura2008} and more realistic modelling, e.g., with local filaments studied by \citet{Bracco2018}.
We intend to extend \hammurabiX\ for further studies of Galactic Faraday rotation, dust emission and free-free absorption by including (where possible) the coupling between the random GMF and the thermal electron and dust distributions implemented in similarly calibrated numeric implementations.

\section*{acknowledgments}
We thank Theo Steininger and Joe Taylor for their contribution in the software development, Sebastian Hutschenreuter for his feedback in using \hammurabiX, and Christopher J.~Anderson for his instructions in using \namaster\,, Dinesh Kandel, Alexandre Lazarian and Dmitri Pogosian for sharing their numerical results.
JW appreciates the pleasant and inspiring discussions with Davide Poletti, Yang Liu, Fran\c{c}ois Boulanger and Anvar Shukurov.
We also thank the constructive comments from the anonymous referee.

The \hammurabiX\ project arose and have received support from the \imagine\footnote{Homepage of the \imagine\ consortium: \url{https://www.astro.ru.nl/imagine/index.html}} meetings hosted by the International Space Science Institute in 2014 and 2015, the Lorentz Center in 2017 and Radboud University in 2019.

The numerical computation is supported by the HPC service and the MHPC program of SISSA.
This work is also partially supported by the National Science Foundation of China (11621303, 11653003, 11773021, 11835009, 11890691), the National Key R\&D Program of China (2018YFA0404601, 2018YFA0404504), the 111 project and the CAS Interdisciplinary Innovation Team (JCTD- 2019-05).

\bibliographystyle{aasjournal}
\bibliography{hammurabiX}

\appendix

\section{synchrotron emission}\label{sec:sync_tech}

In this section, we present the basic mathematical formulae adopted in calculating polarized synchrotron emission and Faraday rotation.
The method is defined not only for analytic modelling of the CRE flux but also for an input grid of dimension $3+1$ imported from external binary files, where the spectral dimension is defined by a logarithmic sampling of electron energy. This matches the output convention in CR transport simulators like Galprop \citep{Strong1998} and DRAGON \citep{Evoli2017}.

\subsection{radiative transfer}

With the CRE differential flux distribution $\Phi(E, \mathbf{r})$,
synchrotron total and polarized emissivities at given observational frequency $\nu$ and spatial position $\mathbf{r}$ read
\begin{eqnarray}\label{eq:emissivity}
    j_\mathrm{tot/pol}(\nu,\mathbf{r}) &=& \frac{1}{4\pi}\int^{E_2}_{E_1} dE \frac{4\pi}{\beta c}\Phi(E,\mathbf{r}) 2\pi P_\mathrm{tot/pol}(\omega) ~,
\end{eqnarray}
where $P_\mathrm{tot/pol}(\omega)$, which represents the emission power from one electron at frequency $\nu = \omega/2\pi$, is calculated \citep{Rybicki1979} through synchrotron functions $F(x) = x\int^\infty_x K_{\frac{5}{3}}(\xi)d\xi$ and $G(x) = xK_{\frac{2}{3}}(x)$ (with $K_{\frac{5}{3}}(x)$ and $K_{\frac{2}{3}}(x)$ known as two of the modified Bessel functions of the second kind) as
\begin{eqnarray}
    P_\mathrm{tot}(\omega) &=& \frac{\sqrt{3}e^3 B_\mathrm{per}}{2\pi m_\mathrm{e}c^2} F(x) ~,\\
    P_\mathrm{pol}(\omega) &=& \frac{\sqrt{3}e^3 B_\mathrm{per}}{2\pi m_\mathrm{e}c^2} G(x) ~,
\end{eqnarray} 
where $e$ is the electron charge, $m_\mathrm{e}$ the electron mass, and $B_\mathrm{per}$ (defined as $|\mathbf{B\times\hat{n}}|$ in Section\,\ref{sec:hamx}) represents the strength of the magnetic field projected in the direction perpendicular to the LoS direction.
Statistically, we assume the synchrotron emission at a given position is isotropic, and so an observer only receives $1/4\pi$ of the emission power, which explains the $1/4\pi$ coefficient in the front of the right-hand-side in equation\,\eqref{eq:emissivity}.
In addition, we place an extra $2\pi$ before $P_\mathrm{tot/pol}(\omega)$ due to the relation $P(\nu) = 2\pi P(\omega)$.
The term $\frac{4\pi}{\beta c}\Phi(E,\mathbf{r})$, with $\beta$ representing the relativistic speed, is actually $N(E,\mathbf{r})$, the CRE differential density.

In practice, the CRE spectral integral can be achieved in two technically different approaches with the same theoretical origin.  
If given numerical CRE flux information $\Phi(E)$ prepared on a discrete grid, the integral equation\,\eqref{eq:emissivity} can be directly evaluated by the numerical integral.
Alternatively we can start with an analytic differential density distribution $N(\gamma, \mathbf{r}) = 4\pi \Phi(E,\mathbf{r}) m_\mathrm{e}c/\beta$, and by doing so the equation\,\eqref{eq:emissivity} reads
\begin{eqnarray}\label{eq:emissivity_alt}
    j_\mathrm{tot/pol}(\nu,\mathbf{r}) &=& \frac{1}{2}\int^{\gamma_2}_{\gamma_1} d\gamma N(\gamma, \mathbf{r}) P_\mathrm{tot/pol}(\omega) ~.
\end{eqnarray}
The reason for keeping equation\,\eqref{eq:emissivity_alt} as an alternative method is to calculate the integral analytically once the CRE spectral index is constant at any given position as illustrated in Section\,\ref{sec:hamx}.
The detailed derivation follows the auxiliary definition of
\begin{eqnarray}
    \omega_c &=& \frac{3}{2}\gamma^2\frac{e B_\mathrm{per}}{m_\mathrm{e} c} ~,\\
    x &=& \frac{\omega}{\omega_c} ~.
\end{eqnarray}
Then by assuming $N(\gamma) = N_0\gamma^{-\alpha}$, equation\,\eqref{eq:emissivity_alt} ends up in the form as
\begin{eqnarray}
    j_\mathrm{tot}(\nu,\mathbf{r}) &=& \frac{\sqrt{3}e^3B_\mathrm{per}N_0}{8\pi m_\mathrm{e}c^2} \left( \frac{4\pi\nu m_\mathrm{e}c}{3eB_\mathrm{per}} \right)^{(1-\alpha)/2} \int dx F(x) x^{(\alpha-3)/2} ~,\\
    j_\mathrm{pol}(\nu,\mathbf{r}) &=& \frac{\sqrt{3}e^3B_\mathrm{per}N_0}{8\pi m_\mathrm{e}c^2} \left( \frac{4\pi\nu m_\mathrm{e}c}{3eB_\mathrm{per}} \right)^{(1-\alpha)/2} \int dx G(x) x^{(\alpha-3)/2} ~,\\
\end{eqnarray}
where the spectral integrals can be analytically calculated by using
\begin{eqnarray}
    \int dx F(x) x^\mu &=& \frac{2^{\mu+1}}{\mu+2}\Gamma\left(\frac{\mu}{2}+\frac{7}{3}\right)\Gamma\left(\frac{\mu}{2}+\frac{2}{3}\right) ~,\\
    \int dx G(x) x^\mu &=& 2^{\mu}\Gamma\left(\frac{\mu}{2}+\frac{4}{3}\right)\Gamma\left(\frac{\mu}{2}+\frac{2}{3}\right) ~.
\end{eqnarray}

\begin{figure}
    \centering
    \includegraphics[width=0.4\textwidth,height=0.4\textwidth]{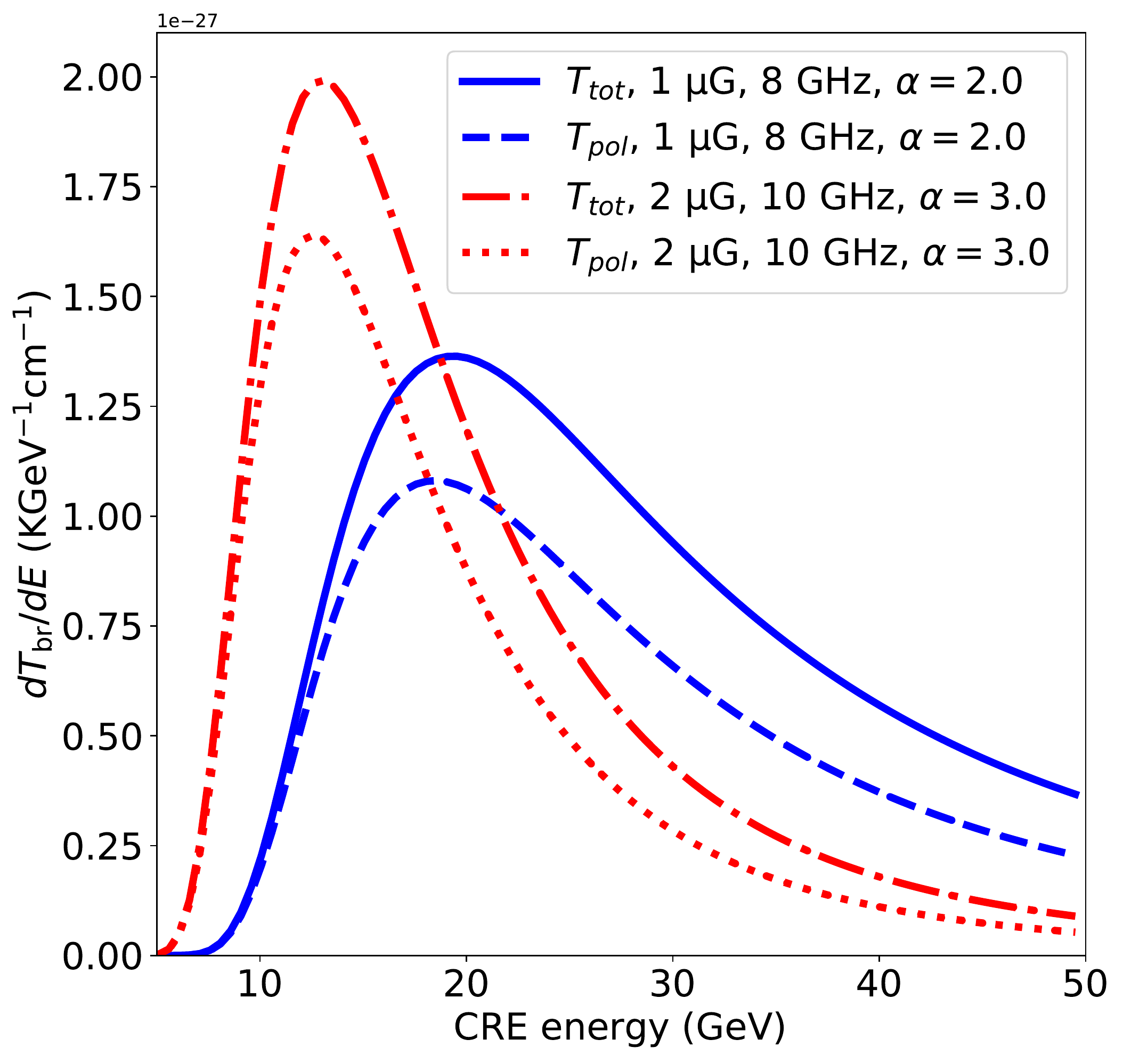}
    \caption{Differential synchrotron total and polarized emissivities ($dj_\mathrm{tot}/dE$ and $dj_\mathrm{pol}/dE$ converted into brightness temperature) of CRE which follows simple power-law spectrum $\propto \gamma^{-\alpha}$. Magnetic field strength and observational frequency are given.}\label{fig:emissivity}
\end{figure}

Figure\,\ref{fig:emissivity} illustrates the dependence of the synchrotron total emissivity $T_\mathrm{tot}$ and polarized emissivity $T_\mathrm{pol}$ on CRE energy, with varying magnetic field strength, observational frequency and CRE spectral shape. 
The peaks in emissivities are inherited from $F(x)$ and $G(x)$, where the dimensionless parameter $x$ is the ratio of observational frequency to CRE gyro-frequency.

In this work, we focus on simulating synchrotron emission at the $\mathrm{GHz}$ level, for which the Galactic environment is optically thin \citep{Rybicki1979,Schlickeiser2002}, and so we ignore both synchrotron self-absorption and free-free absorption.
For readers who might be confused with the synchrotron emissivity calculation formulae presented above, please turn to the \hammurabiX\ wiki page for more technical details.

\subsection{Faraday rotation}

Faraday rotation describes the phenomenological manifestation of the refractive index difference in the polarization directions for photons that propagate through a plasma with an external magnetic field.
For a linearly polarized photon emitted with wavelength $\lambda$ and intrinsic polarization angle $\chi_0$, the observed polarization angle after traversing distance $s_0$ is
\begin{eqnarray}
    \chi &=& \chi_0 + \phi(s_0)\lambda^2 ~,
\end{eqnarray}
where $\phi$, the Faraday depth reads
\begin{eqnarray}
    \phi(s_0) &=& \frac{e^3}{2\pi m_\mathrm{e}^2c^4}\int^{s_0}_0 ds N_\mathrm{e}(s\hat{\mathbf{p}})\mathbf{B}(s\hat{\mathbf{p}})\cdot\hat{\mathbf{p}} ~,
\end{eqnarray}
where $\hat{\mathbf{p}}$ represents photon propagation direction, $N_\mathrm{e}$ represents distribution of thermal electron density.
Note that the IAU convention\footnote{Detailed description for the different IAU and CMB polarization conventions can be found at \url{https://lambda.gsfc.nasa.gov/product/about/pol_convention.cfm}.}
for polarization is adopted in \hammurabiX, which means that the intrinsic synchrotron polarization angle is determined by the polarization ellipse semi-major axis perpendicular to magnetic field orientation.
Under Faraday rotation at a given observational frequency $\nu$, the observed emission accumulates Stokes parameter $dQ$ and $dU$ over a distance $s_0$ by
\begin{eqnarray}
    dQ + idU &=& dI^p_\nu \exp\{2i\chi\} ~,
\end{eqnarray}
where $dI^p_\nu$ represents polarized intensity in radial bin $[s_0,s_0+ds]$.
Though Faraday rotation brings in extra information about the thermal electron (TE) distribution, a relatively high observational frequency is sometimes preferred for studying synchrotron emission, e.g., $30~\mathrm{GHz}$ in this report, to suppress the complicated effects of TE turbulence, which will be addressed in our future studies with \hammurabiX.

\section{precision of random GMF generation}\label{sec:brnd_tech}

In the random GMF generators described in Section\,\ref{sec:gmf}, we are not using three independent FFTs for 3D vector fields.
A straightforward approach to vector field FFT would be carrying out three independent transformations separately.
However, that is expensive in general where the operations are only limited to transforms between real and complex values.
A special speedup design that provides computational efficiency is to compress the three real scalar fields into two complex scalar fields.

Suppose that in the $\xi$-domain we have two complex scalar fields $c_0(\mathbf{\xi})$ and $c_1(\mathbf{\xi})$, which are compressed from three real scalar fields $b_x(\mathbf{\xi})$, $b_y(\mathbf{\xi})$ and $b_z(\mathbf{\xi})$ by defining
\begin{eqnarray}
    c_0(\mathbf{\xi}) &=& b_x(\mathbf{\xi}) + ib_y(\mathbf{\xi}), \\
    c_1(\mathbf{\xi}) &=& b_y(\mathbf{\xi}) + ib_z(\mathbf{\xi}),
\end{eqnarray}
Then mathematically, we know their reciprocal-domain counterparts should be
\begin{eqnarray}
    \tilde{c}_0(\mathbf{\eta}) &=& \tilde{b}_x(\mathbf{\eta}) + i\tilde{b}_y(\mathbf{\eta}), \\
    \tilde{c}_1(\mathbf{\eta}) &=& \tilde{b}_y(\mathbf{\eta}) + i\tilde{b}_z(\mathbf{\eta}).
\end{eqnarray}
Since the transform is done between real and complex fields, complex conjugate symmetry gives a useful property
\begin{eqnarray}
    \tilde{c}^\ast_0(-\mathbf{\eta}) &=& \tilde{b}_x(\mathbf{\eta}) - i\tilde{b}_y(\mathbf{\eta}), \\
    \tilde{c}^\ast_1(-\mathbf{\eta}) &=& \tilde{b}_y(\mathbf{\eta}) - i\tilde{b}_z(\mathbf{\eta}),
\end{eqnarray}
from which we can recover vector fields $\tilde{b}_x(\mathbf{\eta})$, $\tilde{b}_y(\mathbf{\eta})$ and $\tilde{b}_z(\mathbf{\eta})$ in the reciprocal-domain.
This method is applied in both the global and local turbulent GMF generators to reduce the computational cost.

\begin{figure}[htb!]
    \centering
    \includegraphics[width=0.4\textwidth,height=0.4\textwidth]{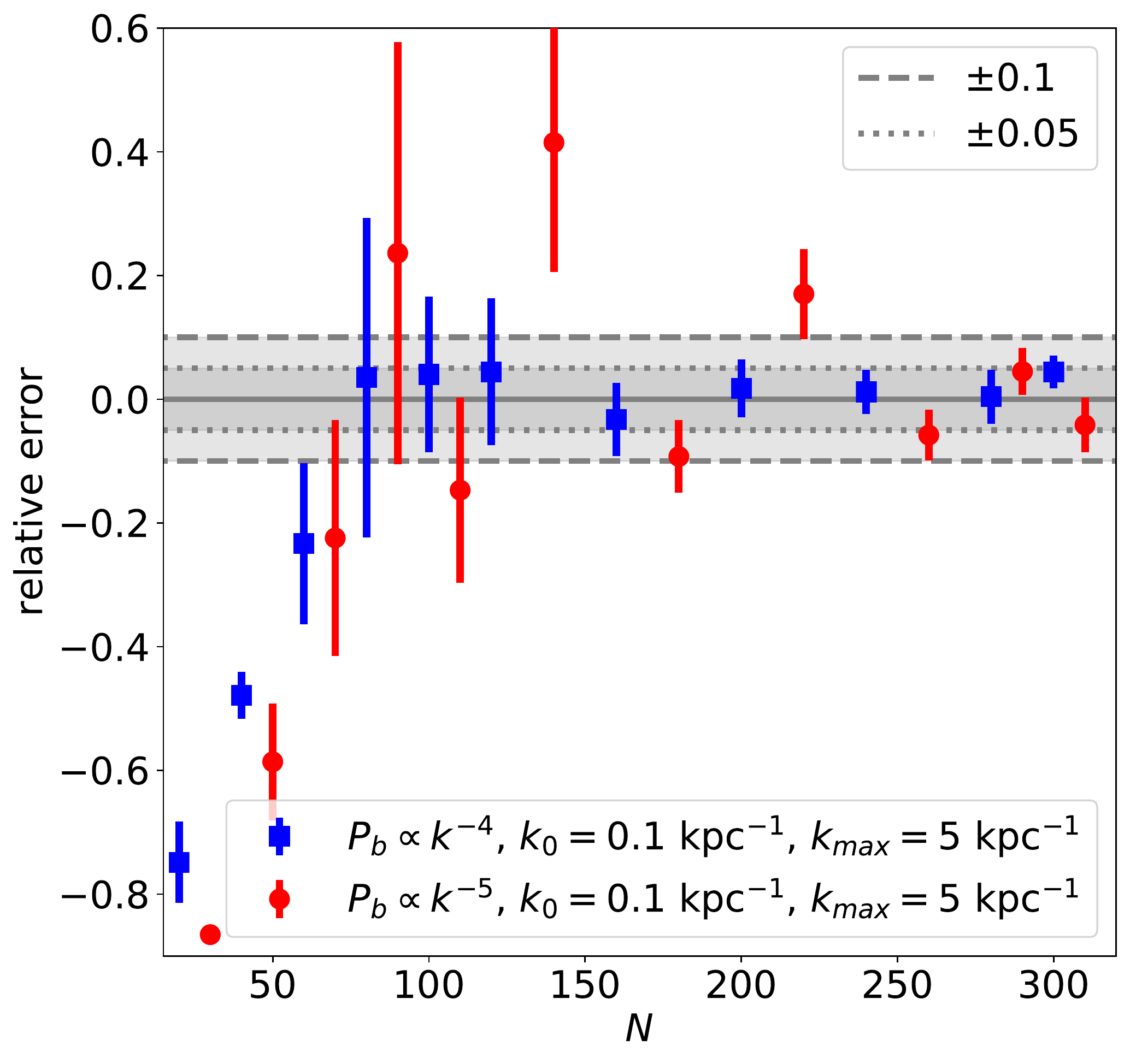}
    \caption{Examples of the relative difference between the theoretical and numerical energy densities in random GMF realizations.
    The numerical energy density of each parameter set is evaluated from an ensemble of field samples.
    A higher precision is achieved with better spatial resolution represented by $N$ (with the simulation box size $L = N/2k_\mathrm{max}$), the number of sample points in each grid dimension.}\label{fig:bvar_precision}
\end{figure}

In the FFTs of both the global and local generators, the numeric field $\mathbf{b(x)}$ is calculated according to its frequency domain counterpart as
\begin{eqnarray}
    \mathbf{b(x)} &=& \sum_{k_x}\sum_{k_y}\sum_{k_z} \mathbf{\tilde{b}(k)} \exp\{2\pi i \mathbf{kx}\} ~.
\end{eqnarray}
Dimensional analysis requires the variance of $\mathbf{\tilde{b}(k)}$ in form
\begin{eqnarray}
    \langle \mathbf{\tilde{b}}_i\mathbf{(k)}\mathbf{\tilde{b}}_j^\ast\mathbf{(k)} \rangle_\mathbf{\tilde{b}} &=& d^3kP_{ij}(\mathbf{k},\theta) ~,
\end{eqnarray}
which in turn satisfies the definition of energy density
\begin{eqnarray}
    E(\mathbf{x}) &=& \frac{\langle \mathbf{b}^2\mathbf{(x)} \rangle_\mathbf{b}}{8\pi} = \int^{k_\mathrm{max}}_0 dk\frac{k^2}{2} \mathrm{Tr}[P_{ij}(k)] ~,
\end{eqnarray}
where $k_\mathrm{max}$ represents the Nyquist frequency.
The precision of the power spectrum as represented on the spatial grid can be visualized by comparing the theoretical and numerical energy densities from field realizations.
As illustrated with examples in Figure\,\ref{fig:bvar_precision}, the convergence towards higher grid resolution demonstrates the correctness of the numeric implementations.

\section{precision of pseudo-$C_\ell$ estimation}\label{sec:cl_tech}

\begin{figure}[htb!]
    \centering
    \includegraphics[width=0.3\textwidth,height=0.3\textwidth]{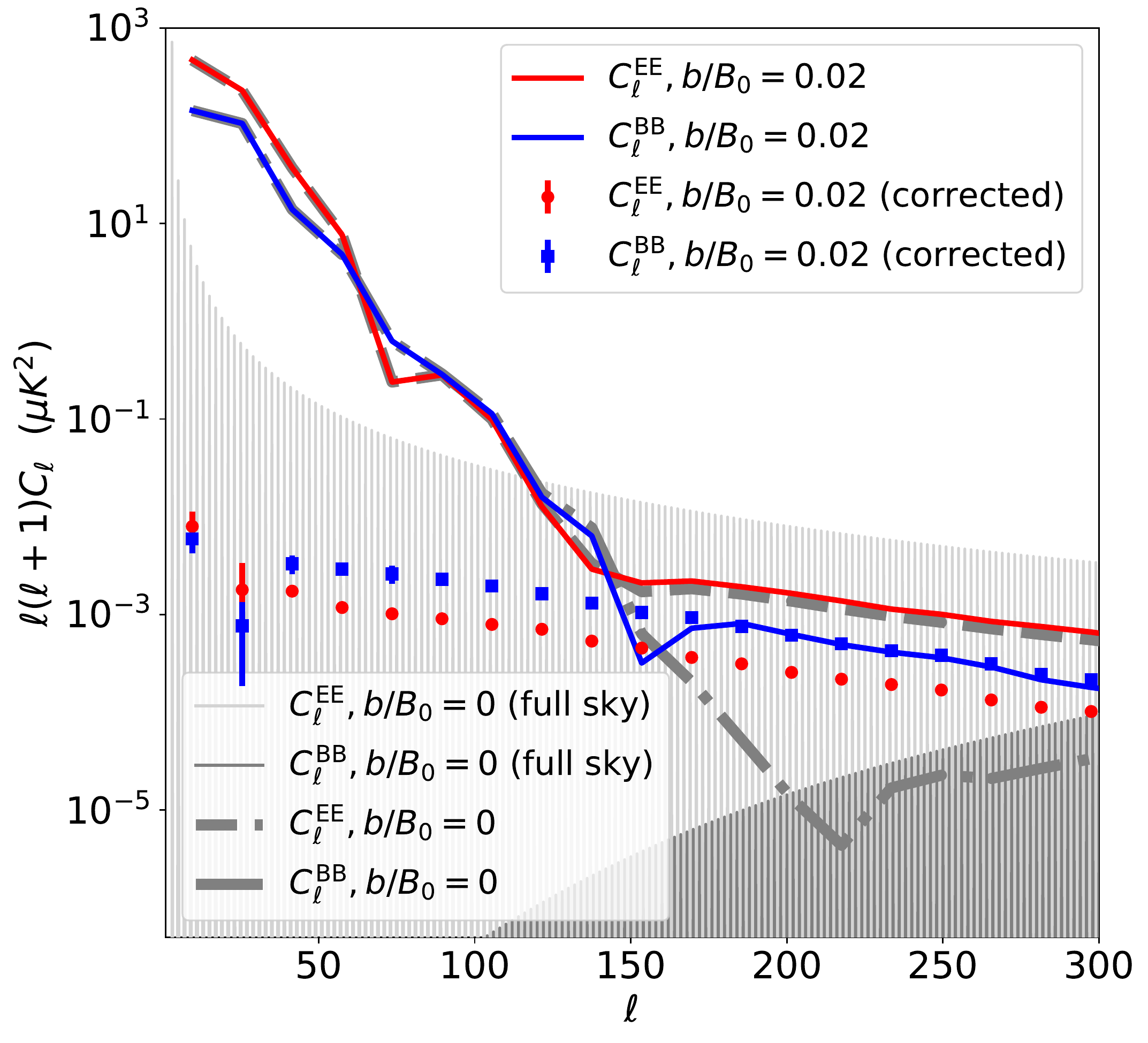}
    \includegraphics[width=0.3\textwidth,height=0.3\textwidth]{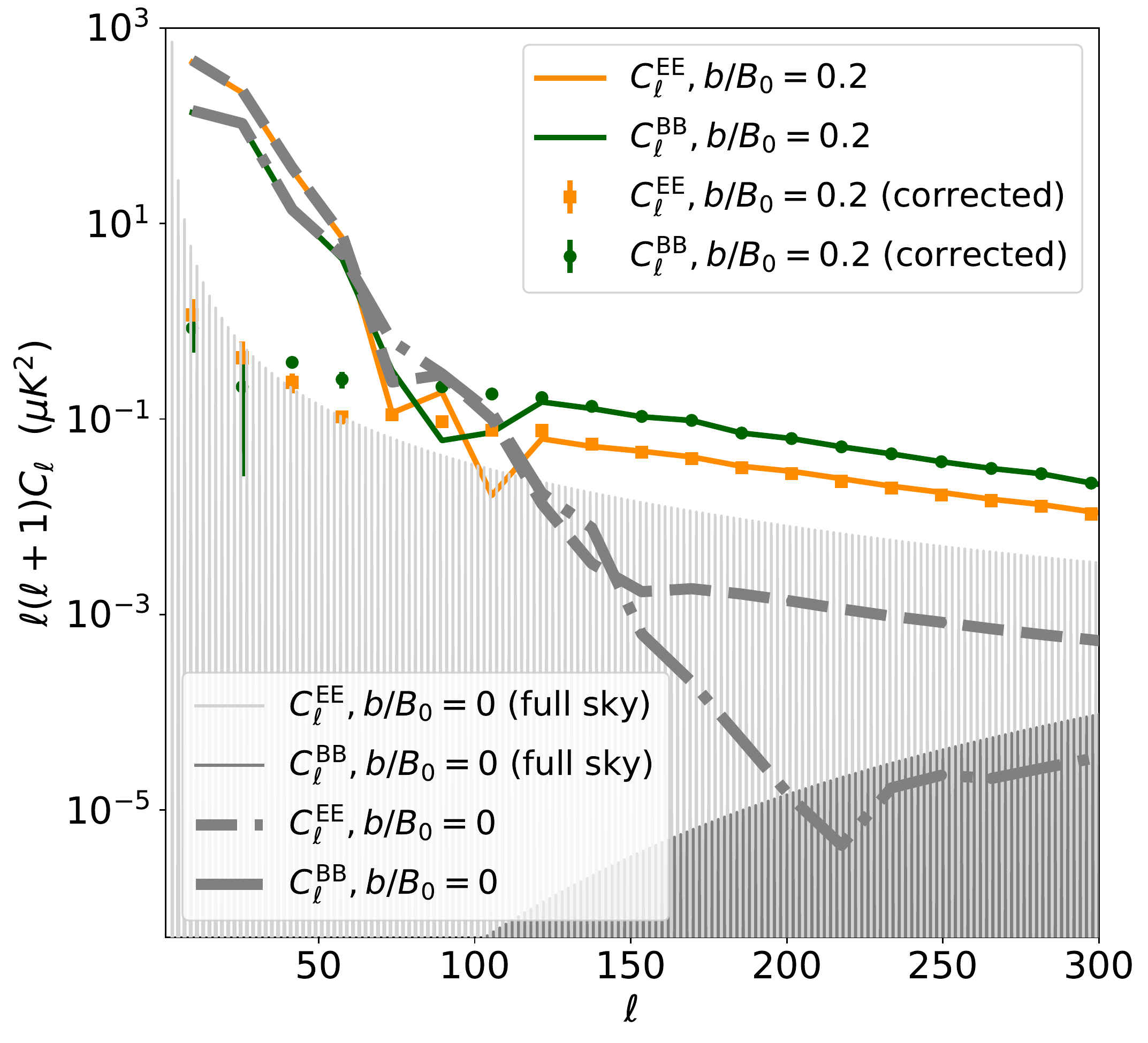}
    \includegraphics[width=0.3\textwidth,height=0.3\textwidth]{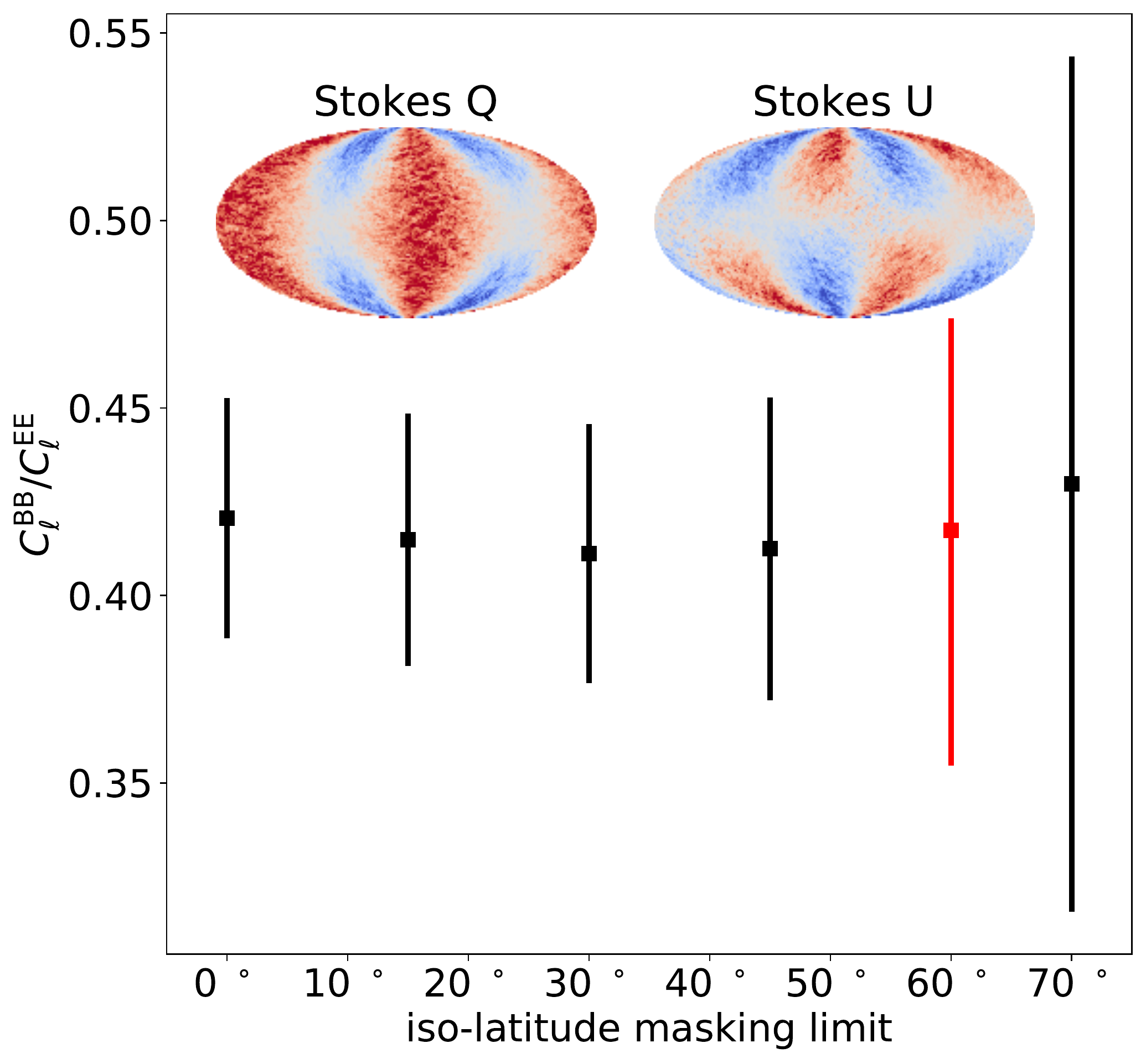}
    \caption{Left and Middle: $C_\ell$s estimated according to global random magnetic fields with $\rho=10.0$ but different strength. The thick grey spectra (dashed and dot-dashed) correspond to the uniform regular magnetic field as defined in Section\,\ref{sec:powerspec}. The light and dark grey shadow solid spectra are from uniform regular magnetic field but estimated from a full-sky map. The shadow areas are actually effects of vanished odd angular modes from the full sky power spectrum estimation. The solid colored curves are estimated pseudo-$C_\ell$ from simulated (partial-sky) outputs, while the square markers with error bars are the estimated pseudo-$C_\ell$ after the regular field contribution being subtracted in the pixel domain. Note that overlap between spectra happens at relative high angular modes.
    Right: One realization of Stokes Q and U maps (according only to the global random magnetic field with $\rho=10.0$) and the corresponding B/E ratio estimated by \namaster\ with various iso-latitude masks. During testing, we find out that setting ten independent realizations in each simulation ensemble is sufficient for getting unbiased estimations.}\label{fig:namaster_demo}
\end{figure}

In this work, the $C_\ell$s are estimated from an ensemble of simulations with the  \namaster\,\footnote{\url{https://github.com/LSSTDESC/NaMaster}} toolkit \citep{Alonso2019}.
Figure\,\ref{fig:namaster_demo} provides some extra information about the pseudo-$C_\ell$ estimation we used.
The iso-latitude masks used here include the one applied in Section\,\ref{sec:Cls}, which corresponds to the $60^\circ$ masking limit in the right panel of Figure\,\ref{fig:namaster_demo}.
To analyze partial-sky observables with the iso-latitude masks with masking limit lower than $70^\circ$ and Gaussian smoothed apodization, we empirically choose band-power binning width $\Delta \ell = 16$ according to the width of the window function.
The regular magnetic field assumed in this work induces a strong large-angular synchrotron polarization.
The symmetry of this synchrotron polarization results in suppression of the odd angular modes in the power spectrum. In the left and middle panels of Figure\,\ref{fig:namaster_demo} the even and odd modes are joined, and the light and dark grey shaded regions represent the E and B mode $C_\ell$s due to the symmetric synchrotron polarization without any masking. 
In presence of a random magnetic field, this suppression of odd harmonics persists at low and intermediate $\ell$ value but goes away at high-$\ell$s.

In case of a partial sky coverage with small sky fraction, like the case considered here, pseudo-$C_\ell$ estimation cannot be done without binning.
However, the suppression of odd harmonics is a complication for pseudo-$C_\ell$ estimators like \namaster.
A pseudo-$C_\ell$ estimate of the symmetric polarization due to the regular magnetic field alone is shown in the grey dashed and dashed-dotted curves in the first two panels of Figure\,\ref{fig:namaster_demo}.
These do not agree with the full sky power spectrum.

\begin{figure}[htb!]
    \centering
    \includegraphics[width=0.4\textwidth,height=0.4\textwidth]{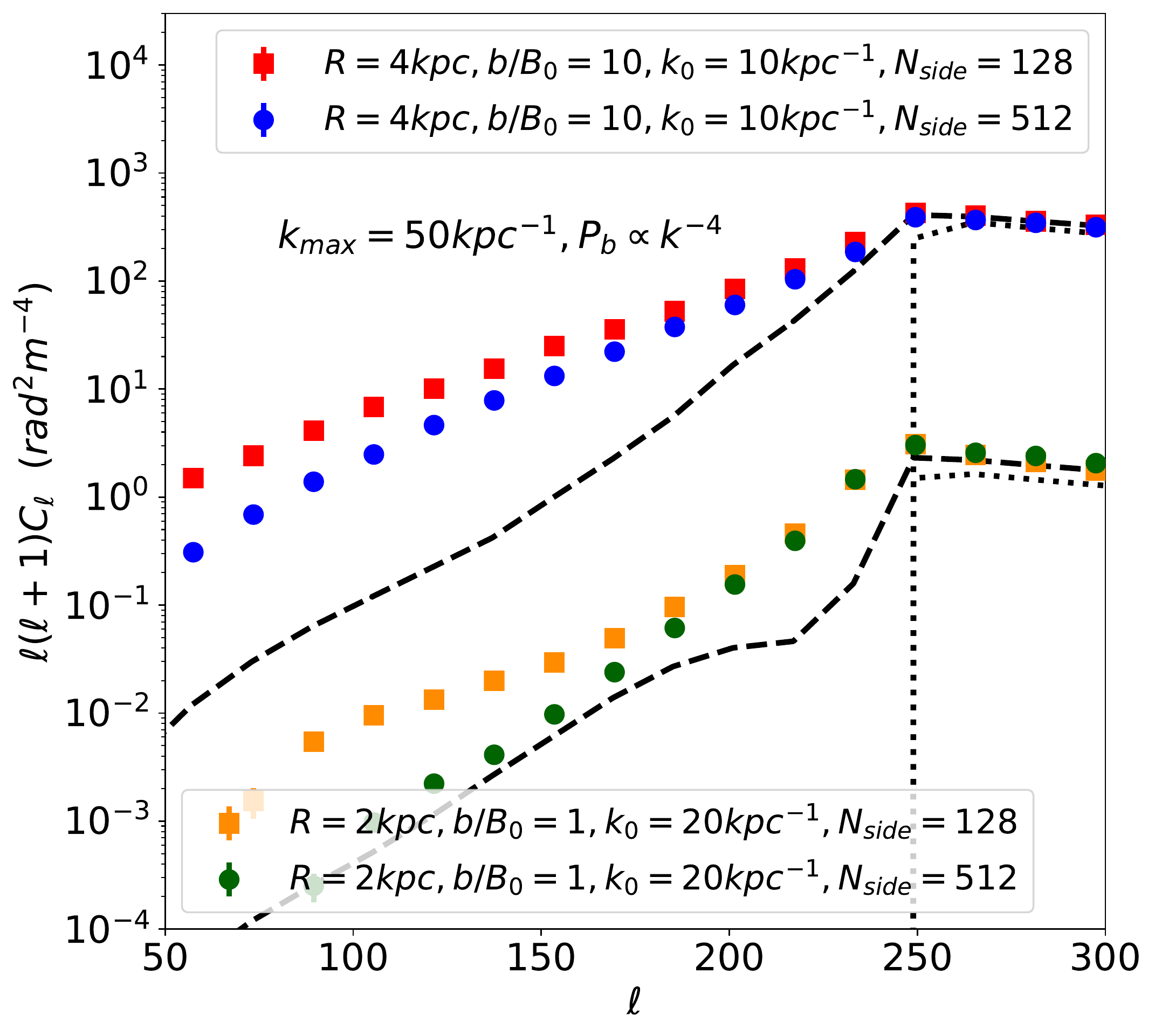}
    \includegraphics[width=0.4\textwidth,height=0.4\textwidth]{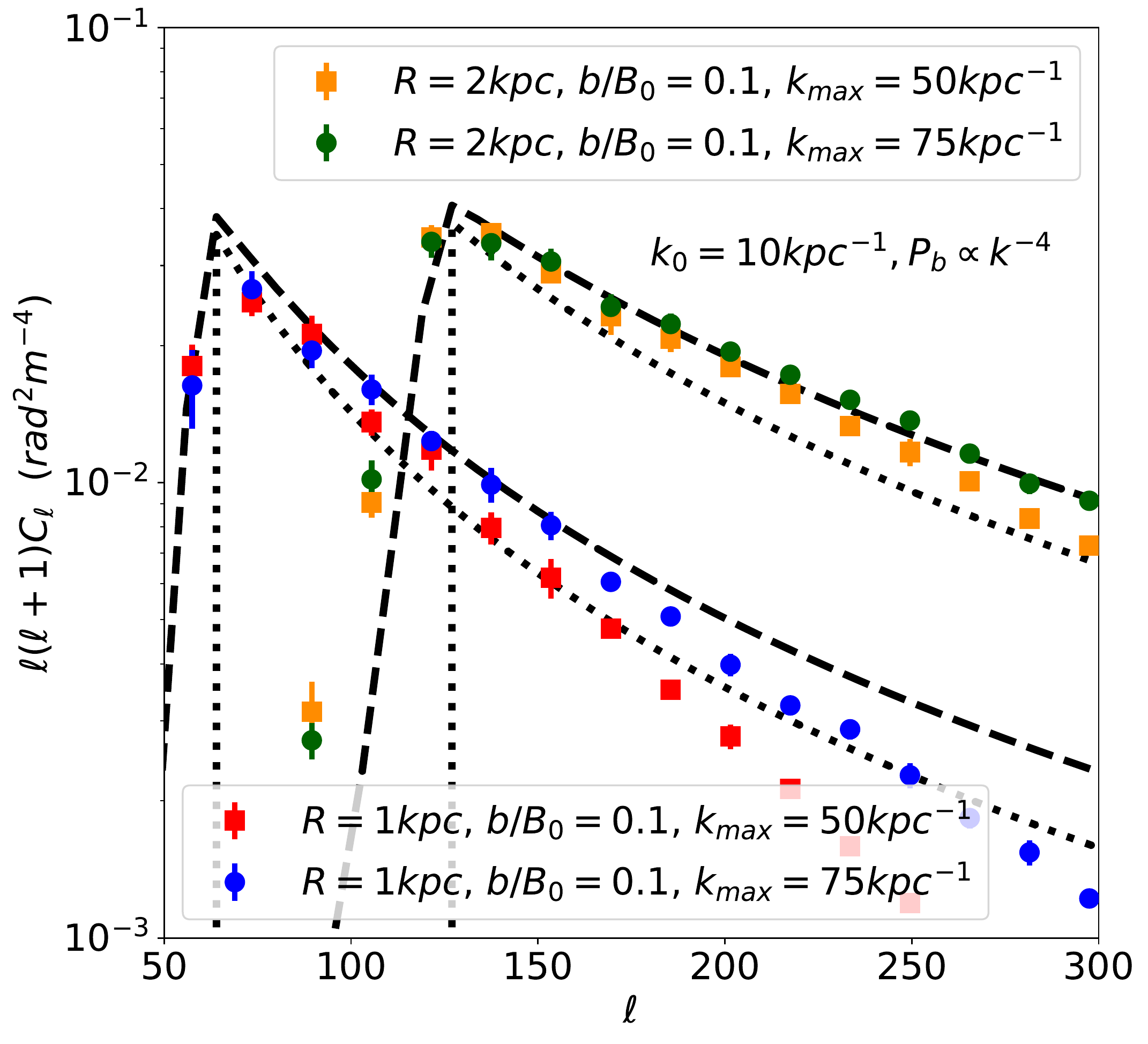}
    \caption{Angular power spectra of Faraday depth estimated on thin shells with central radial distance $R$ and width $\Delta R = 0.1~\mathrm{kpc}$.
    Dotted lines represent estimations made with Limber approximation (equation\,\eqref{eq:clffs}) while dashed lines represent predictions according to numeric integral of spherical Bessel function (equation\,\eqref{eq:clff}).
    Angular power contributed by regular fields has been subtracted.
    }\label{fig:unif_fd_cl_precision}
\end{figure}

The presence of the large-scale symmetry in the polarization presents a critical problem for the pseudo-$C_\ell$ estimation by \namaster, for the total polarization signal produced by the regular and random fields together. 
This may be seen from the solid red/orange and blue/green curves for E and B mode pseudo-$C_\ell$ estimates in the first two panels of Figure\,\ref{fig:namaster_demo}.
These show the identical problem to the plots without a random magnetic field on partial sky. To avoid this problem, in pixel space we subtract the polarization signal produced by the regular magnetic field alone from the total polarization signal. Fortunately, in the illustrative examples, the regular fields are homogeneously defined and so it is feasible and safe to subtract the contribution from the regular magnetic field in the pixel domain. 
We then proceed to use {\namaster} on these `corrected' polarization maps. 
(This is also performed for Figures \ref{fig:unif_gr_becurve}, \ref{fig:unif_lr_becurve} and \ref{fig:unif_fd_cl_precision} as mentioned in the caption.) The pseudo-$C_\ell$ estimates for this `corrected' case is shown in the first two panels of Figure\,\ref{fig:namaster_demo} with red/orange and blue/green data points for E and B mode pseudo-$C_\ell$ estimates respectively. We also show the error bars of reconstruction from 10 independent simulations. 
We restrict our analysis to  $\ell>100$ modes. Note that this correction process only removes the contribution that comes from the regular GMF on its own, i.e., it preserves the polarization signal produced by cross term between the regular and random fields.


We also tried the masking with various latitude limits, and as demonstrated in the right panel of Figure\,\ref{fig:namaster_demo} (where the random magnetic field is generated by the global generator with alignment ratio $\rho=10$), and the B/E ratio estimations are consistent (with larger uncertainty according to smaller sky fraction).

Now we have verified the methods in calculating the synchrotron polarization in Section\,\ref{sec:hamx}, the random field realization in Section\,\ref{sec:gmf}, and the $C_\ell$s in Figure\,\ref{fig:namaster_demo}.
To further confirm the correctness of the simulated results obtained in Section\,\ref{sec:powerspec}, a conceptual verification is necessary.
An analytic approach towards generating the angular power spectrum of tensor fields is not easy and is also beyond our scope.
Alternatively, the shape of the Faraday depth angular power spectrum can be inferred from simplified settings of the fields, which serves as a proper check of the random field realization and the angular modes accumulation in the LoS integral.

To begin with, we adopt the total angular momentum method introduced by \citet{Hu1997,Hu2002}.
Synchrotron polarization $P(r,\hat{\mathbf{n}}) = Q\pm iU$ from a given geocentric position $\mathbf{r} = -r\hat{\mathbf{n}}$ can be expanded in a polarization basis as
\begin{eqnarray}
    P &=& \int \frac{d^3k}{(a\pi)^3}\sum_\ell\sum_{m=-2}^2 [E^{(m)}_\ell \pm B^{(m)}_\ell]\\\nonumber
    &&\times {}_{\pm 2}G^m_\ell(\mathbf{k},r,\hat{\mathbf{n}}) ~,
\end{eqnarray}
where for the spin-2 tensor field the basis reads
\begin{eqnarray}
    {}_{\pm 2}G^m_2 &=& (-i)^\ell \sqrt{\frac{4\pi}{2\ell+1}} {}_{\pm 2}Y^m_2(\hat{\mathbf{n}}) e^{ i\mathbf{kr}} \\\nonumber
    &=& \sum_\ell (-i)^\ell \sqrt{4\pi(2\ell+1)}[\epsilon^{(m)}_\ell(kr)\pm i\beta^{(m)}_\ell(kr)]\\\nonumber
    && \times {}_{\pm 2}Y^m_\ell(\hat{\mathbf{n}}) ~,
\end{eqnarray}
where ${}_{s}Y^m_\ell(\hat{\mathbf{n}})$ is the spherical harmonic function for a spin-$s$ field.
The standard path towards the angular power spectrum E mode $C^{EE}_\ell$ and B mode $C^{BB}_\ell$ starts from interpreting the LoS integral of a target foreground observable with base ${}_{\pm 2}G^m_\ell$ and leads to evaluating
\begin{eqnarray}
    C^{XX}_\ell &=& \frac{4\pi}{(2\ell+1)^2} \int \frac{d^3k d^3q}{(2\pi)^6} e^{i\mathbf{(q-k)x}} \\\nonumber
    &&\times \sum_m \langle X^{(m)\ast}_\ell(\mathbf{k}) X^{(m)}_\ell(\mathbf{q})\rangle ~.
\end{eqnarray}
In the simplest case, we consider only emission sources while ignoring absorption and Faraday rotation, i.e., for a synchrotron polarization tensor $P_\nu(r,\hat{\mathbf{n}})$ at observational frequency $\nu$, 
\begin{eqnarray}
    -\frac{dP_\nu}{dr} &=& \mathcal{S} = j_{pol}e^{2i\chi_0} ~,
\end{eqnarray}
where the basic formulae for polarized emissivity $j_{pol}$ and intrinsic polarization angle $\chi_0$ have been discussed in Appendix\,\ref{sec:sync_tech}.
We would thus expect the integral solution to become
\begin{eqnarray}
    \frac{E^{(m)}_\ell(\mathbf{k})}{2\ell+1} &=& \int dr \frac{{}_{+2}\mathcal{S}^{(m)}_2 + {}_{-2}\mathcal{S}^{(m)}_2}{2} \epsilon^{(m)}_\ell ~,\\
    \frac{B^{(m)}_\ell(\mathbf{k})}{2\ell+1} &=& \int dr \frac{{}_{+2}\mathcal{S}^{(m)}_2 + {}_{-2}\mathcal{S}^{(m)}_2}{2} \beta^{(m)}_\ell ~,
\end{eqnarray}
where the source terms are determined by
\begin{eqnarray}
    \mathcal{S} &=& \int\frac{d^3k}{(2\pi)^3}\sum_m \sum_s ({}_{s}\mathcal{S}^{(m)}_2 {}_{s}G^m_2) ~.
\end{eqnarray}
It is however not trivial (and thus is commonly avoided without further simplification) to analytically bridge the random GMF and its contribution to synchrotron emissivity expanded in a spherical harmonic basis.
Fortunately, Faraday depth is a different story, since the LoS projection of a divergence-free vector field $\mathbf{b}(\mathbf{k})$ can be represented as
\begin{eqnarray}
    \mathbf{b(k)}\cdot\hat{\mathbf{n}} = i\sqrt{\frac{4\pi}{3}}\sum_m b^{(m)}\times{}_{0}Y^m_1(\hat{\mathbf{n}}) ~,
\end{eqnarray}
where the wave-vector $\mathbf{k}$ differs from that in random field realization by a factor of $2\pi$.
(Instead of using the total angular momentum method, a similar approximation to the rotation measure structure function has been carried out by \citet{Xu2016}, which leads to the same conclusion.)
The procedure we take for Faraday depth follows the same method for the Doppler effect handled by \citet{Hu2002}, where the linear perturbation and Limber approximations \citep{LoVerde2008} are key assumptions.
By assuming a uniformly distributed TE field, we isolate the perturbation source of Faraday depth in the vector mode ($m=\pm 1$) which results in the angular power spectrum
\begin{eqnarray}\label{eq:clff}
    C^{FF}_\ell\propto\ell(\ell+1)\int k^2dk P_b(k)\left[\int dr \frac{j_l(kr)}{kr}\right]^2 ~,
\end{eqnarray}
where $P_b$ is power spectrum of random GMF.
By applying Limber approximation (which assumes the typical scale of LoS variation of a perturbed field is much larger than that in the angular direction) we have
\begin{eqnarray}\label{eq:clffs}
    C^{FF}_\ell \propto \int dr P_b\left(\frac{\ell}{r}\right)\frac{1}{r^2} ~,
\end{eqnarray}
which suggests the shape of $C^{FF}_\ell$ is mainly determined by $P_b$.

Figure\,\ref{fig:unif_fd_cl_precision} present a comparison of the simulation precision with respect to the analytic prediction.
For the highest spherical mode $\ell_\mathrm{max}$ in analysis and for a random field grid bin of length $h$, the lower radial limit is roughly set as $R_\mathrm{min} \geq h\ell_\mathrm{max}/\pi$.
Regions closer than $R_\mathrm{min}$ or modes above $\ell_\mathrm{max}$ are greatly affected by the grid interpolation and may affect the pseudo-$C_\ell$ estimation.
The upper radial limit is defined by the simulation size $L$ within which the random GMF is generated, and $R_\mathrm{max} \leq L\ell_\mathrm{min}/\pi$ should be satisfied.
The LoS radius limits discussed here do not influence the conclusions about the B/E ratio but only affecting the precision in estimating $C_\ell$s.
To achieve the highest precision without being distracted by the effects of a multi-shell arrangement, the simulations are done with single shell integrals.
The default simulation and output resolutions are identically set as $N_\mathrm{side}=128$ unless specified.
The random field grid by default is built large enough to host radial integral with LoS depth $R_\mathrm{max}\simeq 4~\mathrm{kpc}$ from the observer with field sampling resolution $h\simeq 3~\mathrm{pc}$ (which means $k_\mathrm{max}\simeq 300~\mathrm{kpc}^{-1}$) and radial resolution $r\simeq 5~\mathrm{pc}$, except that in this appendix we use thin shells with $0.1~\mathrm{kpc}$ thickness and much lower sampling resolution ($k_\mathrm{max}<100~\mathrm{kpc}^{-1}$).
With a sharp cutoff at an injection scale $k_0$ in the random GMF models (by ignoring the inverse cascading), we expect a corresponding break in the angular power spectrum at $\ell_c \sim 2\pi R_\mathrm{max} k_0$.
The break position is well recovered independently of the simulation resolution on each thin LoS shell.
The power in angular modes below and above the break $\ell_c$ is affected differently by the spherical and sampling resolution.
For $\ell < \ell_c$, the angular resolution (characterized by \healpix\ $N_\mathrm{side}$) has a dominant influence, suggesting that a larger angular resolution is necessary for more distant shells to suppress the angular power excess.
While for $\ell > \ell_c$, the missing angular power (particularly for shells closer to the observer) results from insufficient sampling resolution (characterized by the Nyquist frequency $k_\mathrm{max}$) in the random field realization, especially near the observer.
Although the illustrations are prepared with the global random GMF generator, the resolution effects discussed above are generic.
Insufficient angular or Galactic component sampling resolution will result in missing power in the angular power spectra from simulation outputs.
This issue can in principle be handled by using an inhomogeneous grid or adaptively refined mesh with non-equispaced FFT \citep{Keiner2009} for sampling Galactic components (especially the turbulent fields), and also adaptively refined spherical pixelization.
An alternative solution can be nesting sampling grids with different resolutions, but the precision loss on the boundary should be carefully estimated and controlled.
Now with our theoretically verified Faraday depth anisotropy, we can conclude that our numeric realizations of Gaussian random fields are accurate, and thus that the results regarding the B/E ratio obtained from synchrotron emission simulations should be free from numeric defects.

\section{divergence cleaning verification}\label{sec:alternative_tech}

\begin{figure}[htb!]
    \centering
    \includegraphics[width=0.4\textwidth,height=0.4\textwidth]{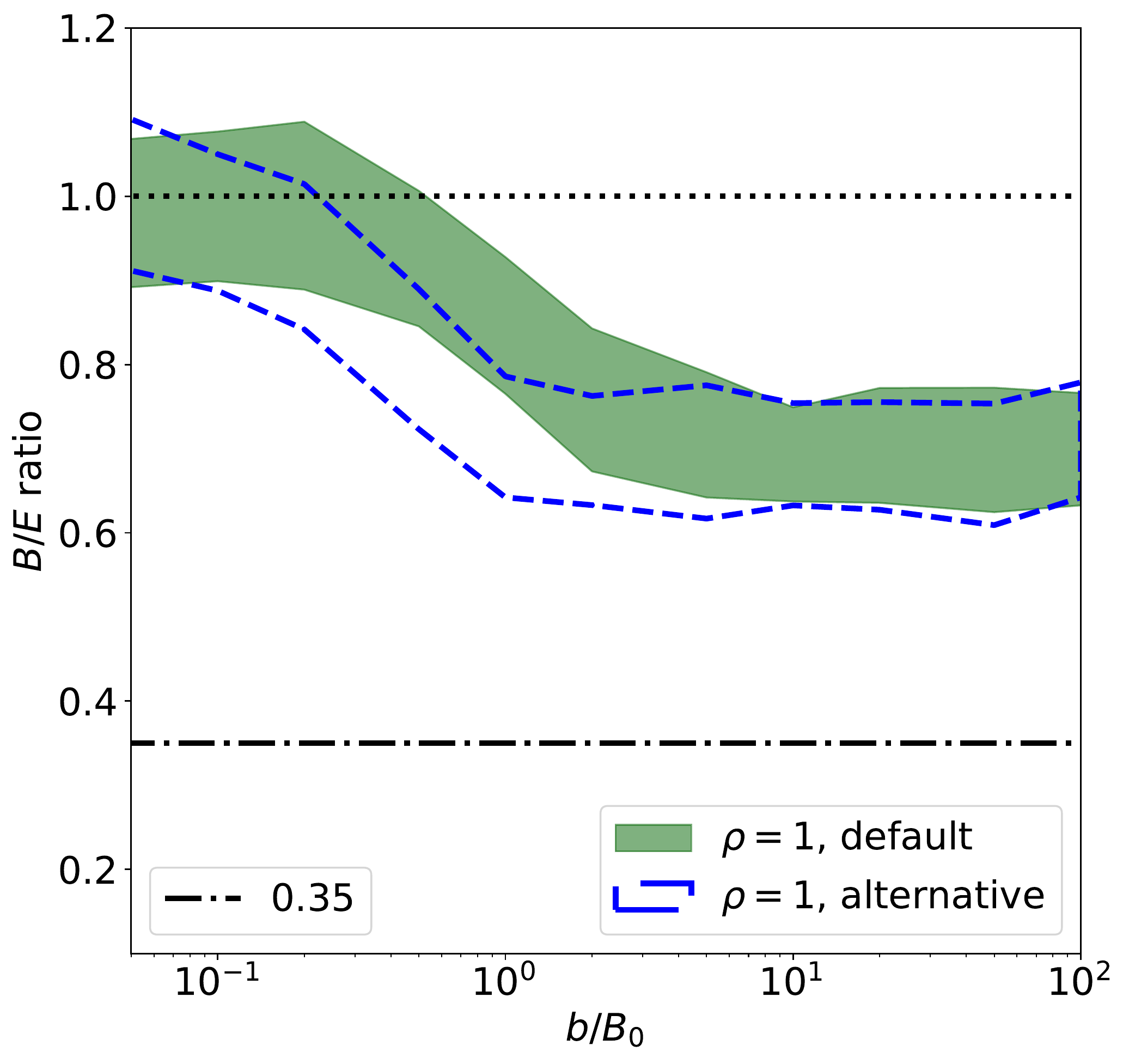}
    \caption{Distribution (16th to 68th percentile) of the $30~\rm{GHz}$ synchrotron emission B/E ratio for $\ell>100$ according to global random GMF with various random field strength.
    The ensemble size is set as ten independent realizations at each sampling position, beyond which we found no significant improvement in the B/E estimation.
    The results marked by ``default'' come from the default algorithm discussed in Section\,\ref{sec:global_generator}, while ``alternative'' indicates random GMF generated from the magnetic potential field realizations.
    The contribution to the angular power spectrum from the regular GMF has been subtracted, which would otherwise dominate the B/E ratio in the perturbative regime ($b \ll B_0$).}\label{fig:unif_gralt_becurve}
\end{figure}

In Section\,\ref{sec:global_generator} we introduced a fast algorithm for generating global random GMFs with divergence cleaning independent from a random sampling of magnetic field vectors in the frequency domain.
To verify the influence of the divergence cleaning on the default global random generator, here we propose an alternative algorithm for generating global Gaussian random GMF by starting with the Gaussian random realizations of the magnetic potential field $\mathbf{A(x)}$.
Knowing a random magnetic field $\mathbf{b(x)}$ can be defined by its potential $\mathbf{A(x)}$, in the frequency domain we have
\begin{eqnarray}
    \mathbf{\tilde{b}}(\mathbf{k}) &=& 2\pi i \mathbf{k}\times \mathbf{\tilde{A}}(\mathbf{k}),
\end{eqnarray}
which ensures $\nabla\times\mathbf{b(x)} = 0$ and so alternatively provides divergence-free random magnetic fields which we can compare to our divergence cleaning using a Gram-Schmidt process.
Note that in this verification, we do not impose any spatial field strength modulation nor orientation alignment, which corresponds to the $\rho=1.0$ case in the default global generator. Figure\,\ref{fig:unif_gralt_becurve} illustrates that the two methods of generating divergence-free random magnetic fields produce equivalent statistical properties of the resulting polarized synchrotron emission.
We have noticed that B/E depends on the ratio between the strength of random and regular magnetic fields (independent of the simulation resolution), as illustrated not only by Figure\,\ref{fig:unif_gralt_becurve} here but also by Figures \ref{fig:unif_gr_becurve} and \ref{fig:unif_lr_becurve}.
This is not predictable by analytic calculations when the random field strength is gradually moving out of the perturbative regime, and it is one of the major advantages and motivations of using \hammurabiX\ for the future studies.

\end{document}